\documentclass[apj,onecolumn]{emulateapj}

\usepackage{amsmath}
\usepackage{empheq}

\shortauthors{LAPI ET AL.}
\shorttitle{ANALYTIC SOLUTIONS FOR GALAXY EVOLUTION II: LATE-TYPE GALAXIES}
\slugcomment{ACCEPTED BY APJ}

\begin{document}

\title{New Analytic Solutions for Galaxy Evolution II: \\Wind Recycling, Galactic Fountains and Late-Type Galaxies}
\author{A. Lapi\altaffilmark{1,2,3,4}, L. Pantoni\altaffilmark{1,3}, L. Boco\altaffilmark{1,2,3}, L. Danese\altaffilmark{1,2}}\altaffiltext{1}{SISSA, Via Bonomea 265, 34136 Trieste, Italy}\altaffiltext{2}{IFPU - Institute for fundamental physics of the Universe, Via Beirut 2, 34014 Trieste, Italy}\altaffiltext{3}{INFN-Sezione di Trieste, via Valerio 2, 34127 Trieste,  Italy}\altaffiltext{4}{INAF-Osservatorio Astronomico di Trieste, via Tiepolo 11, 34131 Trieste, Italy}

\begin{abstract}
We generalize the analytic solutions presented in Pantoni et al. (2019) by including a simple yet effective description of wind recycling and galactic fountains, with the aim of self-consistently investigating the spatially-averaged time evolution of the gas, stellar, metal, and dust content in disc-dominated late-type galaxies (LTGs). Our analytic solutions, when supplemented with
specific prescriptions for parameter setting and with halo accretion rates from $N-$body simulations, can be exploited to reproduce the main statistical relationships followed by local LTGs; these involve, as a function of the stellar mass, the star formation efficiency, the gas mass fraction, the gas/stellar metallicity, the dust mass, the star formation rate, the specific angular momentum, and the overall mass/metal budget. Our analytic solutions allow to easily disentangle the diverse role of the main physical processes ruling galaxy formation in LTGs; in particular, we highlight the crucial relevance of wind recycling and galactic fountains in efficiently refurnishing the gas mass, extending the star-formation timescale, and boosting the metal enrichment in gas and stars. All in all, our analytic solutions constitute a transparent, handy, and fast tool that can provide a basis for improving the (subgrid) physical recipes presently implemented in more sophisticated semi-analytic models and numerical simulations, and can offer a benchmark for interpreting and forecasting current and future spatially-averaged observations of local and higher redshift LTGs.
\end{abstract}

\keywords{galaxies: evolution – galaxies: formation - galaxies: late-type-galaxies}

\section{Introduction}\label{sec|intro}

Understanding the detailed formation path of local late-type galaxies (LTGs) with a prominent disc component is still an open problem in galaxy evolution (see Mo et al. 2010; Silk \& Mamon 2012; Maiolino \& Mannucci 2019; Cimatti et al. 2020). In the last three decades the numerous facets of such a complex issue have been investigated mainly via three methods: hydrodynamical simulations (for a review, see Naab \& Ostriker 2017), semi-analytic models (for a review, see Somerville \& Dave 2015), and analytic frameworks (for a review, see Matteucci 2012).

Numerical simulations will constitute the ultimate approach to address galaxy formation in fine detail; however, despite the recent increase in resolution, many of the relevant physical processes still constitute sub-grid physics, while a detailed exploration of the parameter space is often limited by long computational times. Early cosmological simulations struggled in forming cold, thin and extended stellar disks, due to the overcooling of low-angular momentum baryons at high-redshift (see Katz \& Gunn 1991; Navarro \& Steinmetz 2000; Abadi et al. 2003). Simulations incorporating gas outflows related to stellar feedback mitigated the issue by preventing precocious cooling (see Scannapieco et al. 2008; Governato et al. 2010; Brook et al. 2011), and delaying the star formation history toward the present (see Brook et al. 2012; Stinson et al. 2013), though at the price of yielding outcomes sentitively dependent on the sub-grid recipes and the numerical treatment of the feedback itself (e.g., Scannapieco et al. 2012). Subsequent developments have focused on reproducing the overall galaxy structure, the metal abundance gradients, and the scaling relations among the integrated properties of local disk galaxies (see Guedes et al. 2011; Aumer et al. 2013; Wang et al. 2015;  Colin et al. 2016; Ceverino et al. 2017; Hopkins et al. 2018). Achieving these goals has required to introduce educated star formation thresholds, and to tune high the feedback efficiency. The latter has been possibly considered also in connection with the activity of a central supermassive black holes (see Grand et al. 2017; Valentini et al. 2020). The most recent simulations are starting to fully address the detailed spatial and kinematical structure of spirals and the cycle of multi-phase gas within disk galaxies across cosmic times (see Grand et al. 2019; Pillepich et al. 2019; Vincenzo et al. 2019; Buck et al. 2020).

Semi-analytic models are based on dark matter (DM) merger trees gauged on $N-$body simulations, while the physics inside dark halos is modeled via parametric expressions set on (mainly) local observables. These models are less computationally expensive than hydro simulations
and allow to more clearly disentangle the relative role of
the diverse physical processes; however, the considerable
number of fudge parameters can lead to degenerate
solutions and somewhat limit their predictive power. Early attempts based on simple recipes for cooling and stellar feedback from SN explosions yielded encouraging results in reproducing the properties of local disk galaxies such as sizes, scaling relations, and statistics (see Kauffmann et al. 1993; Lacey \& Cole 1993; Cole et al. 2000;
Baugh et al. 2005). Through the years these basic prescriptions have been progressively refined to describe additional processes and further improve the agreement with observations, even toward high-redshift. Specifically, modern semi-analytic models incorporate: energy feedback from accreting supermassive black holes (see Croton et al. 2006; Somerville et al. 2008; Benson et al. 2012), merger-driven bursts of star formation and dust absorption/emission effects (see Baugh et al. 2005; Cook et al. 2009; Lacey et al. 2016), recycling of blown-out gas via galactic fountains (see Henriques et al. 2015; Croton et al. 2016), metal enrichment in gas and stars (see Cousin et al. 2016; Hirschmann et al. 2016), multiphase treatment of atomic and molecular gas components (see Somerville et al. 2015; Lagos et al. 2018; Baugh et al. 2019), radial structure and gradients (see Stevens et al. 2016; Henriques et al. 2020) and related transport processes of gas and stars (see Forbes et al. 2019).

Frameworks admitting analytic solutions are
necessarily based on approximate and spatially/time-averaged
descriptions of the most relevant astrophysical processes;
however, their transparent, handy, and predictive character
often pays off on some specific issues. Pioneering works were focused on the chemical evolution of the Galaxy, and highlighted the relevance of gas inflow and outflow processes in reproducing the metal abundance of the solar neighborhood (see Schmidt 1963; Talbot \& Arnett 1971; Tinsley 1974; Pagel \& Patchett 1975; Hartwick 1976; Chiosi 1980; Matteucci \& Greggio 1986; Edmunds 1990). Successive developments concerned the mechanisms leading to dust production (see Dwek 1998; Hirashita 2000; Inoue et al. 2003; Zhukovska et al. 2008), the abundance gradients in the Galactic disk (see Chiappini et al. 2001; Naab \& Ostriker 2006; Grisoni et al. 2018), steady-state equilibrium models among star formation, inflows and outflows (see Bouche et al. 2010; Dave et al. 2012; Lilly et al. 2013; Pipino et al. 2014; Feldmann 2015), effects of the initial mass function and of stellar yields models (see Recchi \& Kroupa 2015; Molla et al. 2015), differential and/or selective winds (see Recchi et al. 2008), dichotomy among active and passive galaxies (Spitoni et al. 2017), detailed inside-out growth of galaxy discs with radial mixing (see Andrews et al. 2017; Frankel et al. 2019).

A common conclusion reached by all these different studies and diverse approaches involves the prominent role of feedback processes (for LTGs mainly associated to star formation, i.e. to type-II SN explosions and stellar winds) in originating gas outflows (e.g., White \& Rees 1978; Dekel \& Silk 1986; MacLow \& Ferrara 1999; Murray et al. 2005; Oppenheimer \& Dave 2006; Strickland \& Heckman 2009; Hopkins et al. 2012; Creasey et al. 2013; Heckman \& Thompson 2017; Kim \& Ostriker 2018; Fielding et al. 2018; Hu 2019), that are indeed observed (see Heckman et al. 2000; Pettini et al. 2002; Strickland \& Heckman 2009; Steidel et al. 2010; Arribas et al. 2014; Rubin et al. 2014; Schroetter et al. 2016, 2019). The outflows are found to reduce cooling at high-redshift, to modulate the star-formation efficiency in halos of different masses, to extend the star formation history toward the present, and to drive metal-enriched baryons into the interstellar (ISM) and circumgalactic medium (CGM).

On the other hand, it has also been recognized that, especially in massive LTGs, a fraction of the outflown gas is likely to remain within the DM halo gravitational potential well, and may be able to be recycled and come back at later times in the way of an inward gas flow or of a `galactic fountain' (see Shapiro \& Field 1976; Fall 1979; Bregman 1980).
Through the years the relevance of such a process in the formation and evolution of massive LTGs has been progressively recognized in many respects
(for a review and extended bibliography, see Fraternali 2017). Specifically, it is thought to have profound influence on: chemical evolution in gas and disk stars (see Lacey \& Fall 1985; Pitts \& Tayler 1989; Spitoni et al. 2008, 2009, 2013; Recchi et al. 2008; Calura et al. 2009; Forbes et al. 2014); origin and redistribution of angular momentum and metals (see Mo et al. 1998; van den Bosch 2001; Dutton et al. 2007; Dutton \& van den Bosch 2009; Brook et al. 2012; Stevens et al. 2016, 2018; Grand et al. 2019); kinematics of the cold gas, of the hot corona, and of high-velocity clouds (see Melioli et al. 2008, 2009, 2015; Marinacci et al. 2010, 2011; Fraternali et al. 2015; Li \& Tonnesen 2019); structure of the galaxy neutral hydrogen disk (see Marasco et al. 2012; Stevens \& Brown 2017); origin of abundance gradients (see Fu et al. 2013; Pezzulli \& Fraternali 2016); late fueling and extension of the star formation history toward the present time (see Oppenheimer et al. 2010; Hobbs et al. 2013; Sanchez-Almeida et al. 2014; Tollet et al. 2019).

In Pantoni et al. (2019) we have presented new analytic solutions for the time evolution of the mass, metal, and dust components in star-forming galaxies, with specific focus on the progenitors of early-type galaxies (ETGs). Here we generalize such analytic solutions by including a simple yet effective description of wind recycling and galactic fountains\footnote{To avoid possible misunderstandings, we clarify that the wording `wind recycling and galactic fountain' here means a retention and potential return of the gas and metal mass blown out by stellar feedback into the condensed, star-forming phase within the galaxy.}, with the specific aim of applying them to LTGs.
In this generalized version, our solutions depict the galaxy as an open, one-zone system comprising three interlinked mass components: a reservoir of infalling halo gas able to cool fast, subject to condensation toward the central regions, and refurnished by outflows via wind recycling; cold disk
gas fed by infall and depleted by star formation and stellar
feedback; and stellar mass, partially restituted to the cold phase by
stellar evolution. The metal evolution in cold gas and stellar mass is self-consistently derived from the solutions for the mass components, and
includes the effects of production during star formation, feedback, astration, and fountain. Finally, the dust mass evolution takes also into
account spallation by SN shock waves and accretion of metals onto grain cores. We then supplement the solutions with specific prescriptions for parameter setting and with halo accretion rates from $N-$body simulations, and exploit them to reproduce the main statistical relationships observed in LTGs: the star formation efficiency, the gas mass fraction, the gas/stellar metallicity, the dust mass, the star formation rate, the specific angular momentum, and the mass/metal budget as a function of the stellar mass.

The plan of the paper is as follows. In Sect.~\ref{sec|ansol} we present the new analytic solutions for the time evolution of the gas and stellar
masses (Sect.~\ref{sec|GasStars}), metals (Sect.~\ref{sec|Met}), and dust (Sect.~\ref{sec|dust}); we also show explicitly (Sect.~\ref{sec|nofountain}) that in the limit of no wind recycling our solutions reproduce the expressions by Pantoni et al. (2019). In Sect.~\ref{sec|params} we provide
physical prescriptions to set the parameters entering the analytic solutions for LTGs. In Sect.~\ref{sec|halogrowth} we describe how to include the halo mass growth by merging and accretion from the cosmic web, and how to perform the average of the solution outcomes over different formation redshifts. In Sect.~\ref{sec|results} we compare our results to the available observations concerning the evolution of individual galaxies, the star formation efficiency, the gas mass fraction, the gas metallicity, the stellar metallicity, the main sequence, the dust
mass, the specific angular momentum, and the mass/metal budget.
Finally, in Sect.~\ref{sec|summary} we summarize our approach and main findings.

Throughout this work, we adopt the standard flat $\Lambda$CDM cosmology
(Planck Collaboration 2018) with rounded parameter values: matter density $\Omega_M\approx 0.3$, dark energy density $\Omega_\Lambda\approx 0.7$, baryon density $\Omega_{\rm b}\approx 0.05$, Hubble constant $H_0 = 100\,h$ km s$^{-1}$ Mpc$^{-1}$  with $h\approx 0.7$, and mass variance $\sigma_8\approx 0.8$ on a scale of $8\, h^{-1}$ Mpc. In addition, we use the widely adopted Chabrier (2003, 2005) initial mass function (IMF) with shape $\phi(\log m_\star)\propto \exp[-(\log m_\star-\log 0.2)^2/2\times 0.55^2]$ for $m_\star\la 1\, M_\odot$ and $\phi(\log m_\star)\propto m_\star^{-1.35}$ for $m_\star\ga 1\, M_\odot$, continuously joint at $1\, M_\odot$ and normalized as $\int_{0.1\, M_\odot}^{100\, M_\odot}{\rm d}m_\star\, m_\star\, \phi(m_\star)=1\, M_\odot$. Finally, a value $Z_\odot\approx 0.014$ for the solar metallicity is adopted, corresponding to $12+\log[O/H]_{\odot}=8.69$ (see Allende Prieto et al. 2001).

\section{Analytic solutions for individual LTGs}\label{sec|ansol}

In this section, we present new analytic solutions for galaxy evolution that generalize the ones by Pantoni et al. (2019) by including wind recycling and galactic fountains; these are aimed at describing the spatially-averaged time evolution of the mass, metal, and dust components in individual LTGs. The most relevant expressions are highlighted with a box.

\subsection{Gas and stars}\label{sec|GasStars}

We consider a one-zone description of individual LTGs with three interlinked mass components: the infalling gas mass $M_{\rm inf}$, the cold gas mass $M_{\rm cold}$, and the stellar mass $M_{\star}$; for future reference we define\footnote{In Sect.~\ref{sec|params} it will be clarified that even the infalling gas is cold in the sense it can cool fast, and then it is infalling toward the central region over the dynamical timescale.} $M_{\rm gas}\equiv M_{\rm inf}+M_{\rm cold}$. The evolution of these components as a function of the galactic age $\tau$ is described by the following system of ordinary differential equations (an overdot means differentiation with respect to $\tau$):
\begin{equation}\label{eq|basics}
\left\{
\begin{aligned}
\dot M_{\rm inf} &= -\frac{M_{\rm inf}}{\tau_{\rm cond}}+\alpha_{\rm GF}\, \epsilon_{\rm out}\, \frac{M_{\rm cold}}{\tau_\star}~,\\
\\
\dot M _{\rm cold} &= \frac{M_{\rm inf}}{\tau_{\rm cond}} - (1-\mathcal{R})\,\frac{M_{\rm cold}}{\tau_\star}-\epsilon_{\rm out}\, \frac{M_{\rm cold}}{\tau_\star}~,\\
\\
\dot M_\star &= (1-\mathcal{R})\,\frac{M_{\rm cold}}{\tau_\star}~.\\
\end{aligned}
\right.
\end{equation}
These equations prescribe that the infalling gas mass $M_{\rm inf}$ condenses into the cold gas phase $M_{\rm cold}$ over a characteristic timescale $\tau_{\rm cond}$; then the stellar mass $M_{\rm \star}$ is formed from the cold mass $M_{\rm cold}$ at a rate $M_{\rm cold}/\tau_\star$ over a characteristic timescale $\tau_{\star}$; the cold gas mass is further replenished at a rate $\mathcal{R}\,M_{\rm cold}/\tau_\star$ by stellar recycling, where $\mathcal{R}$ is the return fraction of gaseous material from stellar evolution, and it is removed at a rate $\epsilon_{\rm out}\,M_{\rm cold}/\tau_\star$ by outflows driven from type-II SN explosions and stellar winds, where $\epsilon_{\rm out}$ is the mass loading factor of the outflow; finally, a fraction $\alpha_{\rm GF}$ of this outflowing gas mass returns back to the infalling gas and becomes again available for condensation, in the way of establishing a galactic fountain. In the above Eqs.~(\ref{eq|basics}) the quantity $M_\star$ represents the true relic stellar mass after the loss due to stellar evolution. We adopt an IMF $\phi(m_\star)$ uniform in space and constant in time, and assume the instantaneous mixing (gas is well mixed at anytime) and instantaneous recycling (stars with mass $m_\star\ga 1\, M_\odot$ die as soon as they form, while those with $m_\star\la 1\, M_\odot$ live forever) approximations, so that the recycled fraction (fraction of a stellar population not locked into long-living dark remnants) can be computed as
\begin{equation}
\mathcal{R}\equiv \int_{1\, M_\odot}^{100\, M_\odot}{\rm d}m_\star\, (m_\star-m_{\rm rem})\, \phi(m_\star)
\end{equation}
where $m_{\rm rem}(m_\star)$ is the mass of the remnants; for our fiducial Chabrier (2003, 2005) IMF and the Romano et al. (2010) stellar yield models (see Sect.~\ref{sec|imf}), the recycling fraction amounts to $\mathcal{R}\approx 0.45$. Standard initial conditions for the above system of equations read $M_{\mathrm{inf}}(0)=f_{\rm inf}\, M_{\rm b}$ and $M_{\mathrm{cold}}(0)=M_{\mathrm{\star}}(0)=0$; here $M_{\rm b}=f_{\rm b}\, M_{\rm H}$ is the baryonic mass originally present in the host halo with mass $M_{\rm H}$, while $f_{\rm inf}=M_{\rm inf}/f_{\rm b}\,M_{\rm H}$ is the fraction of such a mass that can effectively cool fast and inflow toward the inner regions of the halo over the timescale $\tau_{\rm cond}$.

The above equations are very similar to those included in many semi-analytic models (SAMs), and in particular in the \texttt{GALFORM} implementation (see Cole et al. 2000; Baugh et al. 2005; Lacey et al. 2016; also Lagos et al. 2018 for a review of different SAMs), though typically these codes solve the equations over infinitesimal timesteps (see Appendix B in Cole et al. 2000), since the halo and infall gas masses are continuously updated due to merging events along DM merger trees. Here instead we provide a global time solution for any finite galactic age, since the latest $N-$body simulations have shown that, when depurated from pseudo-evolution, the mass additions to the baryonic content after formation are minor and occur on rather long timescales so they can be included a posteriori (see discussion in Sect.~\ref{sec|halogrowth}). Note that in Eqs.~(\ref{eq|basics}) we have neglected mechanisms of fountain-driven accretion (see Marinacci et al. 2010, 2011; Marasco et al. 2012; Fraternali et al. 2015; Pezzulli \& Fraternali 2016), envisaging that a significant portion of the CGM that constitutes the hot galactic corona can be induced to condense and fall onto the disk due to the interaction with the ejected gas clouds in the fountain. An attempt to include such an effect by minimally extending our basic analytic framework is presented in the Appendix.

The above system of coupled first order linear differential equations can be solved by writing the first two equations in vectorial form and diagonalising the related coefficients matrix. The resulting eigenvalues read
\begin{equation}\label{eq|eigenvalues}
\left\{
\begin{aligned}
\lambda_+ = \frac{s\gamma+1+\Lambda}{2}\\
\\
\lambda_- = \frac{s\gamma+1-\Lambda}{2}\\
\end{aligned}
\right.
\end{equation}
where $s\equiv \tau_{\rm cond}/\tau_{\rm \star}$, $\gamma=1-\mathcal{R}+\epsilon_{\rm out}$, and $\Lambda=[(s\gamma-1)^2+4\,\alpha_{\rm GF}\,s\, \epsilon_{\rm out}]^{1/2}=\lambda_+-\lambda_-$; with these definitions, note for future reference that $\lambda_+\,\lambda_-=s\,[1-\mathcal{R}+\epsilon_{\rm out}\, (1-\alpha_{\rm GF})]$ and $(1-\lambda_-)\,(\lambda_+-1) = s\,\epsilon_{\rm out}\, \alpha_{\rm GF}$. The solution can be written as
\begin{empheq}[box=\fbox]{align}\label{eq|basicsol}
\nonumber\\
\left\{
\begin{aligned}
M_{\mathrm{inf}}(\tau) &= \frac{f_{\rm inf}\,M_{\mathrm{b}}}{\lambda_+-\lambda_-}\, \left[(\lambda_+-1)\, e^{-\lambda_-\, x}+(1-\lambda_-)\, e^{-\lambda_+\, x}\right]~,\\
\\
M_{\mathrm{cold}}(\tau) &= \frac{f_{\rm inf}\,M_{\mathrm{b}}}{\lambda_+-\lambda_-}\, \left[e^{-\lambda_-\,x}-e^{-\lambda_+\, x}\right]~,\\
\\
M_{\mathrm{\star}}(\tau) &= (1-\mathcal{R})\,\frac{s\,f_{\rm inf}\, M_{\mathrm{b}}}{\lambda_+-\lambda_-}\, \left[\frac{1-e^{-\lambda_-\,x}}{\lambda_-}-\frac{1-e^{-\lambda_+\,x}}{\lambda_+}\right]~,
\end{aligned}
\right.\\
\nonumber
\end{empheq}
where $x\equiv \tau/\tau_{\rm cond}$ is a dimensionless time variable normalized to the condensation timescale. Note that the above solution is physically meaningful (specifically, the cold and stellar masses are non-negative for any $x$) whenever $s\gamma > 1$, which in turn implies that $0<\lambda_-\leq 1\leq \lambda_+$.

It is instructive to examine the initial behavior of the solutions for $\tau\ll \tau_{\rm cond}$, that reads
\begin{equation}
\left\{
\begin{aligned}
M_{\rm inf} &\simeq f_{\rm inf}\, M_{\mathrm{b}}\, \left(1- \frac{\tau}{\tau_{\rm cond}}\right)~,\\
\\
M_{\rm cold} &\simeq f_{\rm inf}\, M_{\mathrm{b}}\, \left(\frac{\tau}{\tau_{\rm cond}}\right)~,\\
\\
M_{\rm \star} &\simeq (1-\mathcal{R})\, \frac{s\, f_{\rm inf}\, M_{\mathrm{b}}}{2}\,\left(\frac{\tau}{\tau_{\rm cond}}\right)^2~;
\end{aligned}
\right.
\end{equation}
the infall and cold gas mass are depleted and enhanced linearly, while the stellar mass rises quadratically since it has to wait the cold gas reservoir to set up. For $\tau\gg \tau_{\rm cond}$ the solutions behave as
\begin{equation}\label{eq|latebasicsol}
\left\{
\begin{aligned}
M_{\mathrm{inf}} &\simeq f_{\rm inf}\, M_{\mathrm{b}}\frac{\lambda_+-1}{\lambda_+-\lambda_-}\, e^{-\lambda_-\,\tau/\tau_{\rm cond}}~,\\
\\
M_{\mathrm{cold}} &\simeq \frac{f_{\rm inf}\, M_{\mathrm{b}}}{\lambda_+-\lambda_-}\, e^{-\lambda_-\,\tau/\tau_{\rm cond}}~,\\
\\
M_{\mathrm{\star}} &\simeq (1-\mathcal{R})\, \frac{s\,f_{\rm inf}\, M_{\mathrm{b}}}{\lambda_+\,\lambda_-}\,\left[1-\frac{\lambda_+}{\lambda_+-\lambda_-}\, e^{-\lambda_-\, \tau/\tau_{\rm cond}}\right]~;
\end{aligned}
\right.
\end{equation}
thus the infall and cold gas masses decline exponentially while the stellar mass converges to the relic value $\bar M_\star\simeq (1-\mathcal{R})\,s f_{\rm inf}\, M_{\rm b}/\lambda_+\lambda_- = (1-\mathcal{R})\,f_{\rm inf}\, M_{\rm b}/[1-\mathcal{R}+\epsilon_{\rm out}\,(1-\alpha_{\rm GF})]$.
Rising for early times and declining at late times, the cold gas mass (and hence the SFR) features a maximum value
\begin{equation}\label{eq|SFRmax}
M_{\rm cold}(\tau_{\rm max}) \simeq \frac{f_{\rm inf}\, M_{\mathrm{b}}}{\lambda_-}\, \left(\frac{\lambda_+}{\lambda_-}\right)^{-\lambda_+/(\lambda_+-\lambda_-)}~.
\end{equation}
occurring at a time
\begin{equation}
\tau_{\rm max} = \tau_{\rm cond}\,\ln\left[\left(\frac{\lambda_+}{\lambda_-}\right)^{1/(\lambda_+-\lambda_-)}\right]~.
\end{equation}

We also stress that the star formation efficiency $f_\star\equiv \bar M_\star/M_{\rm b}$, i.e. the amount of the original baryon content in the halo at formation converted into stars, reads
\begin{empheq}[box=\fbox]{align}\label{eq|fstar}
\nonumber\\
f_{\star} = \frac{1-\mathcal{R}}{1-\mathcal{R}+\epsilon_{\rm out}\,(1-\alpha_{\rm GF})}\, f_{\rm inf}~;\\
\nonumber
\end{empheq}
two remarks are worthwhile. First, all the available (infalling) baryons would be converted into stars ($f_\star\simeq f_{\rm inf}$) in absence of any outflows $\epsilon_{\rm out}\approx 0$, or for a maximally efficient wind recycling $\alpha_{\rm GF}\approx 1$. Second, in presence of fountains the value of the efficiency is generally higher, and its dependence on halo mass, that is encoded mainly in $\epsilon_{\rm out}$ (see Sect.~\ref{sec|params}), is mitigated somewhat; this could be at the origin of the slightly different $f_\star$ measured in halos of the same mass but hosting an LTG or an ETG.

\subsection{Metals}\label{sec|Met}

We now turn to discuss the time evolution of the metallicity in gas and stars, that can be described by the following system of coupled equations:
\begin{equation}\label{eq|Zbasics}
\left\{
\begin{aligned}
{\rm d}_\tau [M_{\rm inf}\,Z_{\rm inf}] &= -\frac{M_{\rm inf}}{\tau_{\rm cond}}\, Z_{\rm inf}+\alpha_{\rm GF}\, \epsilon_{\rm out}\, \frac{M_{\rm cold}}{\tau_\star}\, Z_{\rm cold}~,\\
\\
{\rm d}_\tau [M_{\rm cold}\,Z_{\rm cold}] &= +\frac{M_{\rm inf}}{\tau_{\rm cond}}\, Z_{\rm inf}-\gamma\, \frac{M_{\rm cold}}{\tau_\star}\, Z_{\rm cold}+y_Z\,(1-\mathcal{R})\,\frac{M_{\rm cold}}{\tau_\star}~,\\
\\
{\rm d}_\tau [M_\star\,Z_\star] &= (1-\mathcal{R})\, \frac{M_{\rm cold}}{\tau_\star}\, Z_{\rm cold}~.\\
\end{aligned}
\right.
\end{equation}
The equations above prescribe that the mass of metals in cold gas, $M_{\rm cold}\,Z_{\rm cold}$, evolves because of dilution at a rate $M_{\rm inf}\, Z_{\rm inf}/\tau_{\rm cond}$, instantaneous metal production at a rate $y_Z\, (1-\mathcal{R})\, M_{\rm cold}/\tau_\star$, outflow depletion at a rate $\epsilon_{\rm out}\, M_{\rm cold}\,Z_{\rm cold}/\tau_\star$, and
astration (metal mass locking into stellar remnants) at a rate $(1-\mathcal{R})\,M_{\rm cold}\,Z_{\rm cold}/\tau_\star$; on the other hand, the mass of the metals in the infalling gas varies because of condensation at a rate $M_{\rm inf}\, Z_{\rm inf}/\tau_{\rm cond}$ and of enrichment due to wind recycling at a rate $\alpha_{\rm GF}\, \epsilon_{\rm out}\, M_{\rm cold}\,Z_{\rm cold}/\tau_\star$. We caveat that the above is a basic framework apt for describing analytically the overall metallicity evolution. On the other hand, to follow the evolution of individual elements requires to relax the instantaneous recycling/mixing approximations and to include the chemical enrichment from type-I$a$ SNe, possibly according to a specific delay-time distribution (e.g., Matteucci 2012; for a simple yet still analytic approach, see Pantoni et al. 2019); other complications may arise in the case of differential/selective winds, in which different metals are subject to different feedback efficiencies (e.g., Recchi et al. 2008), or when considering the delayed mixing of gas and metals due to the non-negligible time taken by fountains to orbit around and fall
back onto the galaxy (e.g., Spitoni et al. 2009).

Under the instantaneous mixing and recycling approximation, the metal production yield is given by
\begin{equation}
y_Z\equiv \frac{1}{1-\mathcal{R}}\,\int_{1\, M_\odot}^{100\, M_\odot}{\rm d}m_\star\, m_\star\, p_{Z,\star}\,\phi(m_\star)
\end{equation}
where $p_{Z,\star}$ is the mass-fraction of newly synthesized metals by the star of initial mass $m_\star$; with this definition relative to $1-\mathcal{R}$, the yield $y_Z$ represents the ratio between the mass of heavy elements ejected by a stellar generation and the mass locked up in remnants.

Using Eqs.~(\ref{eq|basics}) the system of the first two equations above can be recasted into the form
\begin{equation}\label{eq|Zgas}
\left\{
\begin{aligned}
\dot Z_{\rm inf} &= \alpha_{\rm GF}\,s\,\epsilon_{\rm out}\, \frac{M_{\rm cold}}{M_{\rm inf}}\, \frac{Z_{\rm cold}-Z_{\rm inf}}{\tau_{\rm cond}}~,\\
\\
\dot Z_{\rm cold} &= -\frac{M_{\rm inf}}{M_{\rm cold}}\, \frac{Z_{\rm cold}-Z_{\rm inf}}{\tau_{\rm cond}}+\frac{s\,y_Z\,(1-\mathcal{R})}{\tau_{\rm cond}}~,
\end{aligned}
\right.
\end{equation}
with initial conditions $Z_{\rm inf}(0)=Z_{\rm cold}(0)=0$. In many previous analytic models, to solve the chemical evolution equation an empirical shape of the SFR is adopted; remarkably, here we instead use the self-consistent solutions for the time evolution of the infalling and cold gas masses, and in particular their ratio
\begin{equation}\label{eq|mratio}
\frac{M_{\rm inf}}{M_{\rm cold}}=\frac{\lambda_+-1+(1-\lambda_-)\, e^{-(\lambda_+-\lambda_-)\,x}}{1-e^{-(\lambda_+-\lambda_-)\,x}}
\end{equation}
that enters Eqs.~(\ref{eq|Zgas}). Plainly, the latter constitute coupled non-linear differential equations, whose solution is generally nontrivial. However, in this particular instance it can be worked out rather easily by taking their difference and obtaining a first order equation for $Z_{\rm cold}-Z_{\rm inf}$, whose solution is trivial; then the latter must be put back on the right hand sides of
Eqs.~(\ref{eq|Zgas}), and a further integration allows to solve analytically for the infalling and cold gas metallicity.

The corresponding analytic solutions read
\begin{empheq}[box=\fbox]{align}\label{eq|Zgassol}
\nonumber\\
\left\{
\begin{aligned}
Z_{\rm inf}(\tau) &= \bar Z_{\rm gas}\left\{(1-\lambda_-)\,x-2\,\frac{1-\lambda_-}{\lambda_+-\lambda_-}\,\left[1-\frac{(\lambda_+-\lambda_-)\,(1+\frac{\lambda_++\lambda_--2}{2}\, x)}{1-\lambda_-+(\lambda_+-1)\,e^{(\lambda_+-\lambda_-)\,x}}\right]\right\}~,\\
\\
Z_{\rm cold}(\tau) &= \bar Z_{\rm gas}\left\{(1-\lambda_-)\,x+\frac{\lambda_++\lambda_--2}{\lambda_+-\lambda_-}\,\left[1-\frac{(\lambda_+-\lambda_-)\,x}{e^{(\lambda_+-\lambda_-)\,x}-1}\right]\right\}~;\\
\end{aligned}
\right.\\
\nonumber
\end{empheq}
here
\begin{equation}\label{eq|Zgasave}
\bar Z_{\rm gas} = \frac{s\, y_Z\, (1-\mathcal{R})}{\lambda_+-\lambda_-}~
\end{equation}
represents the asymptotic value for $\tau\gg \tau_{\rm cond}$ of the difference $Z_{\rm cold}-Z_{\rm inf}$, which is finite since individually $Z_{\rm cold}$ and $Z_{\rm inf}$ diverge linearly in the same manner. Note that such a divergence is not an issue because the masses in metals $M\, Z$ for both the infall and the cold gas components are always finite and exponentially suppressed at late-times.

It is also worth looking at the initial behavior of the gas metallicities for $\tau\ll \tau_{\rm cond}$, that reads
\begin{equation}
\left\{
\begin{aligned}
Z_{\rm inf} &\simeq \frac{s\,y_Z\,(1-\mathcal{R})}{6}\, s\,\epsilon_{\rm out}\,\alpha_{\rm GF}\, \left(\frac{\tau}{\tau_{\rm cond}}\right)^3~,\\
\\
Z_{\rm cold} &\simeq \frac{s\,y_Z\,(1-\mathcal{R})}{2}\, \frac{\tau}{\tau_{\rm cond}}~;
\end{aligned}
\right.
\end{equation}
the cold gas metallicity increases almost linearly with galactic age, while the infalling gas metallicity evolution is cubic; this is because at early times the cold gas is rapidly enriched by direct production from stars, while the infalling gas has to wait the galactic fountain process to bring some metals from the cold gas into it.

The last of Eqs.~(\ref{eq|Zbasics}) implies that the metallicity $Z_{\rm \star}$ in the stellar component is the average of the cold gas one over the star formation history:
\begin{equation}\label{eq|Zstar_def}
Z_{\rm \star}(\tau) = \frac{1}{M_{\rm \star}(\tau)}\, \int_0^{\tau}{\rm d}\tau'~Z_{\rm cold}(\tau')\,\dot M_{\rm \star}(\tau')~.
\end{equation}
Using the self-consistent solutions for $Z_{\rm cold}$ and $M_{\rm cold}$, one obtains
\begin{empheq}[box=\fbox]{align}\label{eq|Zstar_inst}
\begin{aligned}
\\
& Z_{\rm \star}  = \bar Z_\star\,\left\{1-\cfrac{\lambda_+\,\lambda_-}{(\lambda_+-\lambda_-)\,[\lambda_+\,(1-e^{-\lambda_-\,x})-\lambda_-\,(1-\,e^{-\lambda_+\,x})]}\,\left[\cfrac{2\,\lambda_+\lambda_--\lambda_+-\lambda_-}{\lambda_+-\lambda_-} \right. \right.\\
\\
& \left.\left.  \times\left(e^{-\lambda_-\,x}-e^{-\lambda_+\,x}\right)+(1-\lambda_-)\,\lambda_+\,x\,e^{-\lambda_-\,x}-(\lambda_+-1)\,\lambda_-\,x\,e^{-\lambda_+\,x}\cfrac{}{}\right]\right\}~,\\
\\
\end{aligned}
\end{empheq}
where the limiting value for $\tau\gg \tau_{\rm cond}$ writes
\begin{equation}\label{eq|Zstarave}
\bar Z_{\rm \star} = \frac{s\,y_Z\,(1-\mathcal{R})}{\lambda_+\,\lambda_-} = \frac{y_Z\,(1-\mathcal{R})}{1-\mathcal{R}+\epsilon_{\rm out}\,(1-\alpha_{\rm GF})}~;
\end{equation}
it is seen that our analytic solutions predict that the asymptotic stellar metallicity $\bar Z_\star\simeq \bar Z_{\rm gas}\, (\lambda_+-\lambda_-)/\lambda_+\lambda_-$ is not equal, but rather somewhat lower, than the gas one. The early-time behavior of $Z_\star$ for $\tau\ll \tau_{\rm cond}$  reads
\begin{equation}
Z_{\rm \star}\simeq \frac{s\,y_Z\, (1-\mathcal{R})}{3}\, \frac{\tau}{\tau_{\rm cond}}~,
\end{equation}
so that initially $Z_{\star}(\tau)\simeq 2\, Z_{\rm cold}(\tau)/3$, i.e., the stellar and cold gas metallicity evolve in parallel.

We conclude by stressing that Eqs.~(\ref{eq|fstar}) and (\ref{eq|Zstarave}) implies a direct connection among the infalling gas fraction $f_{\rm inf}$, the asymptotic stellar metallicity $\bar Z_\star$ and the star formation efficiency $f_\star$, in the form
\begin{empheq}[box=\fbox]{align}\label{eq|finf}
\nonumber\\
f_{\rm inf} = \frac{y_Z\,f_\star}{\bar Z_\star}~;\\
\nonumber
\end{empheq}
this has indeed been used by Shi et al. (2017) and Lapi et al. (2018a)
to infer the infall fraction $f_{\rm inf}$ via observations of $f_\star$ (from photometric/dynamical modeling and weak lensing data) and $\bar Z_\star$ (from stellar archeology).

\subsection{Dust}\label{sec|dust}

We now turn to describe the global evolution of the dust mass and dust-to-gas mass ratio, taking up the basic analytic modeling adopted by many previous studies (e.g., Dwek 1998; Hirashita 2000; Edmunds 2001; Inoue et al. 2003; Asano et al. 2013; Feldmann 2015; Mancini et al. 2015). Specifically, we assume dust to consist of two interlinked components, namely, a refractory \emph{core} and a volatile \emph{mantle}, subject to the evolution equations\footnote{In principle, a negative term describing gas mass and metal locking into dust should be added on the right hand side of the differential equations for ${\rm d}_\tau M_{\rm cold}$ and ${\rm d}_\tau [M_{\rm cold}\, Z_{\rm cold}]$; however, these terms are usually neglected since they are proportional to the dust mass and this is always a small fraction of the gas one.}
\begin{equation}\label{eq|dust}
\left\{
\begin{aligned}
{\rm d}_\tau [M_{\rm cold}\, D_{\rm core}] & =-\gamma\,\dot{M}_{\rm \star}\, D_{\rm core}-\kappa_{\rm SN}\,\dot{M}_{\rm \star}\, D_{\rm core}+{y_D}\,(1-\mathcal{R})\, \dot{M}_{\rm \star}~,\\
\\
{\rm d}_\tau [M_{\rm cold}\, D_{\rm mantle}] &=-\gamma\,\dot{M}_{\rm \star}\, D_{\rm mantle}-\kappa_{\rm SN}\,\dot{M}_{\star}\, D_{\rm mantle}+ \epsilon_{\rm acc}\, \dot{M}_{\rm \star}\, D_{\rm core}\,(Z_{\rm cold}-D_{\rm mantle})~.
\end{aligned}
\right.
\end{equation}
The first equation prescribes that the evolution of the mass in grain cores $M_{\rm cold}\,D_{\rm core}$ results from the competition of various processes: production due to stellar evolution at a rate $y_D\,(1-\mathcal{R})\,\dot{M}_{\rm \star}$ with an average yield $y_D$; astration by starformation and ejection from galactic outflows, that combine in the rate term $-\gamma\, \dot M_{\rm \star}\, D_{\rm core}$; dust sputtering, spallation and destruction via SN shockwaves at a rate $\kappa_{\rm SN}\, \dot{M}_{\rm \star}\, D_{\rm core}$ with a strength parameter $\kappa_{\rm SN}$. The second equation describes the evolution of the mass in dust mantles, which differs from the previous one for the production term: mantle growth is assumed to be driven by accretion of metals onto pre-existing grain cores at a rate $\epsilon_{\rm acc}\, \dot M_{\rm \star}\, D_{\rm core}\, (Z-D_{\rm mantle})$ with an efficiency $\epsilon_{\rm acc}$.

Eqs.~(\ref{eq|dust}) can be recast in terms of the dust-to-gas mass ratios:
\begin{equation}\label{eq|Dbasics}
\left\{
\begin{aligned}
\dot{D}_{\rm core}&= -\frac{D_{\rm core}}{\tau_{\rm cond}}\, \left[s\,\kappa_{\rm SN}+\frac{M_{\rm inf}}{M_{\rm cold}}\right]+\frac{y_D\, (1-\mathcal{R})\,s}{\tau_{\rm cond}}~;\\
 \\
\dot{D}_{\rm mantle}&= -\frac{D_{\rm mantle}}{\tau_{\rm cond}}\, \left[s\,\kappa_{\rm SN}+s\,\epsilon_{\rm acc}\, D_{\rm core}+\frac{M_{\rm inf}}{M_{\rm cold}}\right]+\frac{s\, \epsilon_{\rm acc}\, D_{\rm core}}{\tau_{\rm cond}}\, Z_{\rm cold}~.
\end{aligned}
\right.
\end{equation}
with initial conditions $D_{\rm core}(0)=D_{\rm mantle}(0)=0$; in the above both $M_{\rm inf}/M_{\rm cold}$ and $Z_{\rm cold}$ depend on $\tau$ (or $x$) as expressed by Eqs.~(\ref{eq|mratio}) and (\ref{eq|dust}), respectively.

The corresponding analytic solution for grain cores is
\begin{empheq}[box=\fbox]{align}\label{eq|Dcore}
\nonumber\\
D_{\rm core} = \bar D_{\rm core}\, \left[1-\frac{\lambda_+-\lambda_-}{e^{(\lambda_+-\lambda_-)\, x}-1}\, \frac{1-e^{-(s\, \kappa_{\rm SN}+\lambda_--1)\, x}}{s\, \kappa_{\rm SN}+\lambda_--1}\right]~,\\
\nonumber
\end{empheq}
where the asymptotic value for $\tau\gg \tau_{\rm cond}$ reads
\begin{equation}\label{eq|Dcoreave}
\bar D_{\rm core} = \frac{s\,y_D\, (1-\mathcal{R})}{s\,\kappa_{\rm SN}+\lambda_+-1}~.
\end{equation}

In solving the equation for the mantle, we assume the core fraction $D_{\rm core}$ to be fixed at its asymptotic value $\bar D_{\rm core}\ll \bar Z_{\rm cold}$, since from Eq.~(\ref{eq|Dcore}) this is seen to be attained quite rapidly after a time $\tau\ga \tau_{\rm cond}/s\, \kappa_{\rm SN}$. We then obtain
\begin{empheq}[box=\fbox]{align}
\begin{aligned}
\\
D_{\rm mantle} &= \bar D_{\rm mantle}\, \left\{\frac{1}{d_{\pm}}\, \frac{(1-\lambda_-)\,x}{1-e^{-(\lambda_+-\lambda_-)\, x}}+1-\frac{1}{ d_{\pm}}\,\frac{s\tilde\epsilon}{s\tilde\epsilon+\lambda_--1}  \,\frac{(\lambda_+-1)\, x}{e^{(\lambda_+-\lambda_-)\, x}-1}\right.\\
\\
&\times \left.\left[1+\frac{\lambda_+-1}{s\tilde\epsilon}\,
\left(1-\frac{(\lambda_+-\lambda_-)^2}{(\lambda_+-1)^2}\,\frac{s\,\tilde \epsilon+\lambda_++\lambda_--2}{s\,\tilde \epsilon+\lambda_+-1}\,\frac{1-e^{-(s\,\tilde\epsilon+\lambda_--1)\, x}}{(s\,\tilde \epsilon+\lambda_--1)\, x}\right)\right]\right\}\\
\\
\end{aligned}
\end{empheq}
where $\tilde\epsilon\equiv \kappa_{\rm SN}+\epsilon_{\rm acc}\,\bar D_{\rm core}$ and we have defined the two auxiliary quantities
\begin{equation}\label{eq|Dmantleave}
\bar D_{\rm mantle} = \frac{s\,\epsilon_{\rm acc}\, \bar D_{\rm core}\, \bar Z_{\rm gas}}{s\,\tilde\epsilon+\lambda_+-1}\, d_{\pm}~
\end{equation}
and
\begin{equation}\label{eq|Dmantleaux}
d_{\pm}\equiv \frac{\lambda_++\lambda_--2}{\lambda_+-\lambda_-}-\frac{1-\lambda_-}{s\, \tilde \epsilon+\lambda_+-1}~.
\end{equation}
We stress that if the accretion process is very efficient, Eq.~(\ref{eq|Dmantleave}) implies that the final dust fraction tends to behave like the gas metallicity.

The early-time behavior for $\tau\ll \tau_{\rm cond}$ writes
\begin{equation}
\left\{
\begin{aligned}
D_{\rm core}(\tau) &\simeq \frac{s\,y_D\,(1-\mathcal{R})}{2}\,\frac{\tau}{ \tau_{\rm cond}}~,\\
\\
D_{\rm mantle}(\tau) &\simeq \frac{s^3}{6}\, \frac{\epsilon_{\rm acc}\, y_D\, y_Z\,(1-\mathcal{R})^2}{s\, (\gamma+\kappa_{\rm SN})-1}\, \left(\frac{\tau}{\tau_{\rm cond}}\right)^2~,
\end{aligned}
\right.
\end{equation}
so that the mantle component overwhelms the core one soon after dust production has started. Note that the second equation above is strictly valid for times $\tau_{\rm cond}/s\kappa_{\rm SN}\la \tau\la \tau_{\rm cond}$ when $D_{\rm core}$ has already saturated to its asymptotic value $\bar D_{\rm core}$.

It is worth stressing that total dust production and enrichment (core plus mantle) are in general very rapid with respect to the condensation timescale, of order a few $10^{-1}\, \tau_{\rm cond}$.

\subsection{Limit of no galactic fountain}\label{sec|nofountain}

When wind recycling and galactic fountains are negligible, it is straightforward to demonstrate that the solutions above converge to the ones already presented in Pantoni et al. (2019). In fact, posing $\alpha_{\rm GF}\simeq 0$ in Eqs.~(\ref{eq|eigenvalues}) implies $\lambda_+\simeq s\,\gamma$ and $\lambda_-\simeq 1$. Thus the solutions for the gas and stellar masses become
\begin{equation}
\left\{
\begin{aligned}
M_{\mathrm{inf}}(\tau) &= f_{\rm inf}\, M_{\mathrm{b}}\, e^{-x}~,\\
\\
M_{\mathrm{cold}}(\tau) &= \frac{f_{\rm inf}\, M_{\mathrm{b}}}{s\,\gamma-1}\, \left[e^{-x}-e^{-s\, \gamma\, x}\right]~,\\
\\
M_{\mathrm{\star}}(\tau) &= (1-\mathcal{R})\,\frac{s\,f_{\rm inf}\, M_{\mathrm{b}}}{s\,\gamma-1}\, \left[1-e^{-x}-\frac{1}{s\,\gamma}\,\left(1-e^{-s\, \gamma\, x}\right)\right]~,
\end{aligned}
\right.
\end{equation}
and those for the gas and stellar metallicities turn into
\begin{equation}
\left\{
\begin{aligned}
Z_{\rm inf} &\simeq 0~,\\
\\
Z_{\rm cold} &\simeq \bar Z_{\rm gas}\, \left[1-\frac{(s\,\gamma-1)\, x}{e^{(s\, \gamma-1)\, x}-1}\right]~,\\
\\
Z_\star &\simeq \bar Z_{\rm \star}\, \left[1-\frac{s\, \gamma}{s\, \gamma-1}\, \frac{e^{-x}-e^{-s\, \gamma\, x}\, [1+(s\, \gamma-1)\, x]}{s\gamma-1+e^{-s\, \gamma\, x}-s\gamma\,e^{-x}}\right]~,
\end{aligned}
\right.
\end{equation}
where now the asymptotic values read $\bar Z_{\rm gas} = s\, y_Z\, (1-\mathcal{R})/(s\gamma-1)$ and $\bar Z_\star = y_Z\, (1-\mathcal{R})/\gamma$.

As for the dust evolution, consider that the auxiliary quantity $d_{\pm}$ defined in Eq.~(\ref{eq|Dmantleaux}) tends to $1$ and hence the solutions become
\begin{equation}
\left\{
\begin{aligned}
D_{\rm core} &\simeq \bar D_{\rm core}\, \left[1-\frac{s\, \gamma-1}{e^{(s\, \gamma-1)\, x}-1}\, \frac{1-e^{-s\, \kappa_{\rm SN}\, x}}{s\, \kappa_{\rm SN}}\right]~,\\
\\
D_{\rm mantle} &\simeq \bar D_{\rm mantle}\, \left\{1-\frac{(s\, \gamma-1)\, x}{e^{(s\, \gamma-1)\, x}-1}\, \left[1+\frac{s\, \gamma-1}{ s\, \tilde\epsilon}\, \left(1-\frac{1-e^{-s\,\tilde\epsilon\, x}}{s\,\tilde \epsilon\, x}\right)\right]\right\}~,
\end{aligned}
\right.
\end{equation}
in terms of the quantity $\tilde\epsilon\equiv \kappa_{\rm SN}+\epsilon_{\rm acc}\, \bar D_{\rm core}$ and of the asymptotic values $\bar D_{\rm core}\simeq s\,y_D\, (1-\mathcal{R})/[s\,(\gamma+\kappa_{\rm SN})-1]$ and $\bar D_{\rm mantle}\simeq s\,\epsilon_{\rm acc}\, \bar D_{\rm core}\, \bar Z_{\rm cold}/[s\,(\gamma+\tilde\epsilon)-1]$.

All in all, these are the solutions provided by Pantoni et al. (2019).

\section{Parameter setting}\label{sec|params}

We now set the parameters entering the previous analytic solutions, taking up the physical arguments laid down in Lapi et al. (2018a) and Pantoni et al. (2019; see their Sect.~3), to which we refer the reader for details. Here we also add some specific prescriptions required to describe the formation and evolution of LTGs.

\subsubsection{Infall fraction and condensation timescale}\label{sec|infall}

Given a halo of mass $M_{\rm H}$ at redshift $z$, its virial radius and virial circular velocity are approximately given by $R_{\rm H}\approx 260\, M_{\rm H,12}^{1/3}$ $\,E_{z}^{-1/3}$ kpc and $v_{c, \rm H} \approx 130\, M_{\rm H,12}^{1/3}\, E_{z}^{1/6}$ km s$^{-1}$ in terms of the redshift dependent factor $E_{z}=\Omega_\Lambda+\Omega_M\,(1+z)^3$ and of the normalized halo mass $M_{\rm H,12}=M_{\rm H}/10^{12}\, M_\odot$.
The halo mass profile is assumed to follow a standard NFW shape $M_{\rm H}(<r)\propto \ln(1+c\,x)-c\,x/(1+c\,x)$ with $x\equiv r/R_{\rm H}$; the concentration $c\approx c_{\rm form}\, (1+{z_{\rm form}})/(1+z)$ applies in terms of the value $c_{\rm form}\approx 4$ at the formation redshift $z_{\rm form}$ (see Bullock et al. 2001; Zhao et al. 2003, 2009; Prada et al. 2012; Correa et al. 2015; Klypin et al. 2016; Child et al. 2018; Diemer \& Joyce 2019).

\begin{figure}
\centering
\includegraphics[width=\textwidth]{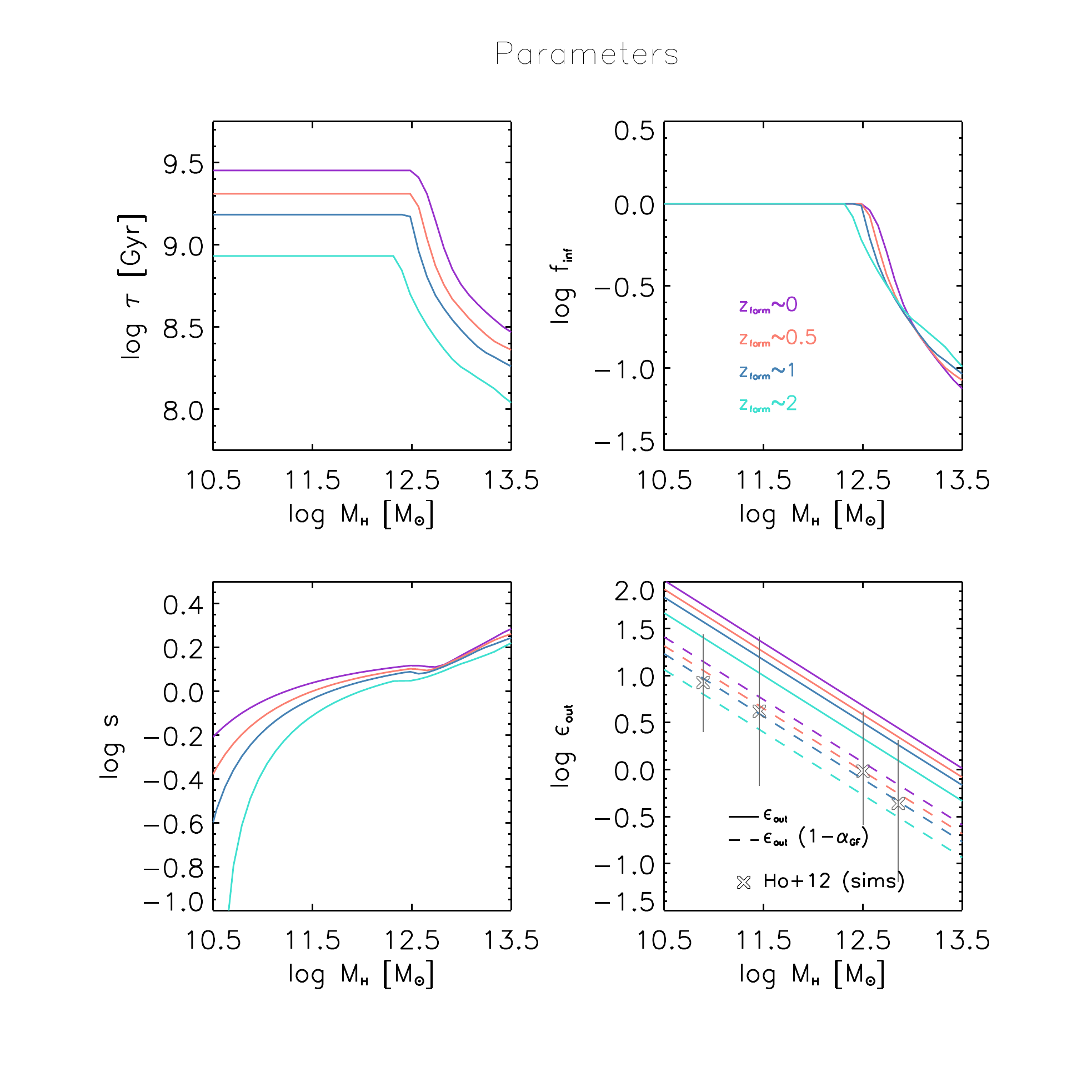}
\caption{Main parameters entering the analytic solutions relevant for LTGs as a function of host halo mass $M_{\rm H}$ and formation redshift $z_{\rm form}$. Top left panel: condensation timescale $\tau_{\rm cond}$. Top right panel: infall fraction $f_{\rm inf}$. Bottom left panel: ratio
$s\equiv \tau_{\rm cond}/\tau_\star$ of the condensation to the star formation timescale. Bottom right panel: mass loading factor of the outflows from stellar feedback $\epsilon_{\rm out}$ for energy-driven winds;
solid lines refer to $\epsilon_{\rm out}$ and dashed lines to $\epsilon_{\rm out}\, (1-\alpha_{\rm GF})$ with $\alpha_{\rm GF}=0.75$, while crosses illustrate the outcomes from hydrodynamic simulations of stellar feedback by Hopkins et al. (2012). In all panels, the colors refer to different formation redshifts: $z_{\rm form}=0$ (purple), $0.5$ (orange), $1$ (blue), and $2$ (cyan).}\label{fig|params}
\end{figure}

A fraction $f_{\rm inf}$ of the available baryons $f_{\rm b}\, M_{\rm H}$, initially located with an infall radius $R_{\rm inf}\simeq f_{\rm inf}\, R_{\rm H}$, is able to cool fast and fall in toward the central region of the galaxy where star formation takes place. We set $f_{\rm inf}$ by requiring that the dynamical time
\begin{equation}
t_{\rm dyn}(R_{\rm inf})\simeq\frac{\pi}{2}\sqrt{\frac{R_{\rm inf}^3}{G\,M_{\rm tot}(<R_{\rm inf})}}\approx 3\times 10^9\,f_{\rm inf}\,E_{z}^{-1/2}~\rm{yr}~,
\end{equation}
and the cooling time
\begin{equation}
t_{\rm cool}(R_{\rm inf})\simeq
\frac{3\,k_{\rm B}T}{\mu\, \mathcal{C}\,n(R_{\rm inf})\,\Lambda(T,Z)}\approx 4.5\times 10^{9}\, f_{\rm inf}^2\, \cfrac{M_{\rm H,12}^{2/3}\, E_z^{-2/3}}{\Lambda_{-23}[T(M_{\rm H},z),Z]}~\rm{yr}~
\end{equation}
match, so that within $R_{\rm inf}$ the condition $t_{\rm cool}\lesssim t_{\rm dyn}$ is met (see Rees \& Ostriker 1977; White \& Rees 1978) and the gas is allowed to condense over the timescale $\tau_{\rm cond}\simeq t_{\rm dyn}(R_{\rm inf})$. In the above expressions we have approximated $M_{\rm tot}(<R_{\rm inf})\simeq M_{\rm H}(<R_{\rm inf})\simeq f_{\rm inf}\, M_{\rm H}$ and used
$T\simeq 0.5\,\mu\,m_p\,v_{c,\rm H}^2/k_{\rm B}\approx 6\times 10^5\, M_{\rm H,12}^{2/3}\,E_{z}^{1/3}$ K as the virial temperature, $\mu\approx 0.6$ as the mean molecular weight, $n(R_{\rm inf})\approx 3 \times 10^{-5}\, f_{\rm inf}^{-2}\,E_{z}$ cm$^{-3}$ as the gas density at $R_{\rm inf}$ for an isothermal spherical distribution, $\mathcal{C}\sim 10$ as the clumping factor (see Iliev et al. 2007; Pawlik et al. 2009; Finlator et al. 2012; Shull et al. 2012), and $\Lambda_{-23}(T,Z)\equiv \Lambda/10^{-23}$ K cm$^3$ s$^{-1}$ as the cooling function in cgs units dependent on temperature and metallicity (e.g., Sutherland \& Dopita 1993). We plot $f_{\rm inf}$ and $\tau_{\rm cond}$ as a function of the halo mass and formation redshift in Fig.~\ref{fig|params}. The parameter $f_{\rm inf}$ is essentially unity for halo masses $M_{\rm H}\la$ a few $10^{12}\, M_\odot$, while for larger masses it drops to low values because cooling becomes progressively inefficient and prevents condensation toward the central regions; the dependence on formation redshift is negligible. As to $\tau_{\rm cond}$, it scales with halo mass similarly to $f_{\rm inf}$, while the constant value for $M_{\rm H}\la$ a few $10^{12}\, M_\odot$ depends on  redshift approximately as $(1+z)^{-3/2}$, reflecting the increased density of the ambient medium at earlier cosmic epochs.

\subsubsection{Molecular gas fraction and star-formation timescale}\label{sec|molgas}

The infalling gas rotates, being endowed with the specific angular momentum
\begin{equation}\label{eq|jhalo}
j_{\rm inf} \equiv f_{\rm inf}\,j_{\rm H}\approx 1670\, f_{\rm inf}\,\lambda_{0.035}\,M_{\rm H,12}^{2/3}\, E_{z}^{-1/6}~{\rm km~s^{-1}~kpc}~,
\end{equation}
where $j_{\rm H} \equiv \sqrt{2}\, \lambda\, R_{\rm H}\, v_{c, \rm H}$ is the halo one, in terms of the spin parameter $\lambda_{0.035}\equiv \lambda/0.035$ normalized to the average value
$0.035$ found in numerical simulations (see Barnes \& Efstathiou 1987; Bullock et al. 2001; Macci\'o et al. 2007; Shi et al. 2017; Zjupa \& Springel 2017). The inflow of the gas from $R_{\rm inf}$ toward the central region can proceed until the radius $R_{\rm rot}$ where the rotational support balance the gravitational pull; the condition $G\, M_{\rm tot}(<R_{\rm rot})/R_{\rm rot}^2=j_{\rm inf}^2/R_{\rm rot}^3$ yields
\begin{equation}\label{eq|Rrot}
R_{\rm rot}\simeq \frac{j_{\rm inf}^2}{G\, f_{\rm inf}\, f_{\rm b}\, M_{\rm H}}\, \delta_{\rm rot}\approx 3.5\, \lambda_{0.035}^2\,f_{\rm inf}\,M_{\rm H,12}^{1/3}\, E_{z}^{-1/3}~~{\rm kpc}~,
\end{equation}
where $\delta_{\rm rot}\equiv f_{\rm inf}\, f_{\rm b}\, M_{\rm H}/M_{\rm tot}(<R_{\rm rot})\approx 0.9$ is the mass contrast within $R_{\rm rot}$ (see Lapi et al. 2018a for details). In passing, we note that if $R_{\rm rot}$ is identified with the disk scale-length, then the ratio between the half-mass radius $R_e\approx 1.68\, R_{\rm rot}$ and the halo size $R_{\rm H}$ takes on values $R_{\rm e}/R_{\rm H}\approx 0.023\, \lambda_{0.035}^2\, f_{\rm inf}$, in remarkable agreement with the recent finding by Zanisi et al. (2020; see also Kravtsov et al. 2013). The gas there will form star on a timescale (see Elmegreen et al. 2005; Krumholz et al. 2012 and references therein)
\begin{equation}\label{eq|tdyn_Rrot}
\tau_\star\simeq 50\, \cfrac{t_{\rm dyn}(R_{\rm rot})}{f_{\rm H_2}}\approx 6\times 10^8\,\lambda_{0.035}^{3}\, f_{\rm H_2}^{-1}\,f_{\rm inf}\, E_{z}^{-1/2}~~{\rm yr}~.
\end{equation}
where $t_{\rm dyn}(R_{\rm rot})\simeq (\pi/2)\sqrt{R_{\rm rot}^3/G\, M_{\rm tot}(<R_{\rm rot})}$ is the dynamical time at $R_{\rm rot}$, and  $f_{\rm H_2}\equiv M_{\rm H_2}/M_{\rm cold}$ is the molecular gas fraction; such a timescale corresponds to a star formation law $\dot M_\star \propto f_{\rm H_2}\, M_{\rm cold}$ in terms of the last Eqs.~(\ref{eq|basics}) as observed in local starforming galaxies (e.g., Schruba et al. 2011).

Basing on empirical evidences, models aiming at a spatially-resolved description of LTGs employ a dependence of the molecular gas fraction on the local disk pressure, on gas metallicity and/or on the interstellar radiation field (e.g. Blitz \& Rosolowski 2006; Leroy et al. 2008; Krumholz 2013; Fu et al. 2013; Gnedin \& Draine 2014; see also Lagos et al. 2018 for a more detailed review of these assumptions). Here we take up the suggestion put forward by Obreschkow et al. (2016) and Murugeshan et al. (2019), which is based on the observational finding that the atomic and molecular gas fraction scale almost linearly with $j/M$, which in turn is proportional to the (mass-weighted) Toomre (1964) parameter $Q\equiv \sqrt{2}\,\Omega\,\sigma/\pi\, G \Sigma$. In fact, the latter involves the angular rotation velocity $\Omega\equiv v/R\simeq j/R^2$, the intrinsic velocity dispersion $\sigma$ of the gas related to turbulent motions (note that the interstellar medium is likely to become multi-phase after infall, see Braun \& Schmidt 2012), and the average gas surface density $\Sigma\simeq M_{\rm gas}(<R)/\pi R^2$. Combining these definitions yields the scaling
\begin{equation}\label{eq|Toomre}
Q \simeq \sqrt{2}\, j_{\rm inf}\, \sigma/G f_{\rm inf}\, f_{\rm b}\, M_{\rm H}\approx 0.03\, \sigma_{10\,(1+z)}\,\lambda_{0.035}\,M_{\rm H,12}^{-1/3}\, E_{z}^{-1/6}\,(1+z)~;
\end{equation}
the gas intrinsic velocity dispersion $\sigma_{10\,(1+z)}\equiv \sigma/10\, (1+z)$ km s$^{-1}$ has been normalized to a fiducial value of $10$ km s$^{-1}$ as measured in local starforming galaxies, and then taken to increase as $(1+z)$ with the redshift as observed out to $z\lesssim 3$ (see Wisnioski et al. 2015; Saintonge et al. 2017; Johnson et al. 2018).

\begin{figure}
\centering
\includegraphics[width=\textwidth]{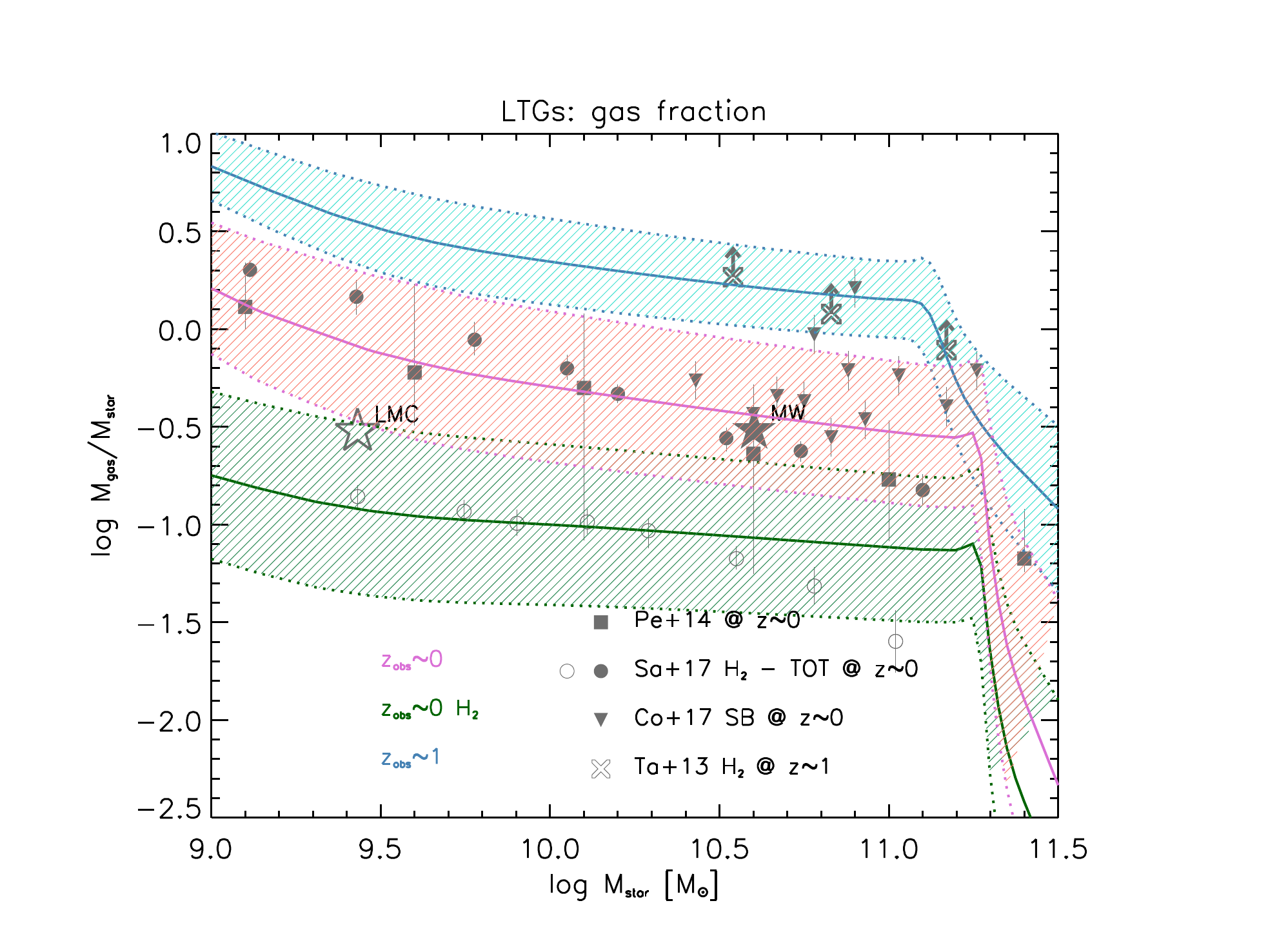}
\caption{Gas fraction $f_{\rm gas}\equiv M_{\rm gas}/M_\star$ as a function of the stellar mass $M_\star$ at observation redshifts $z_{\rm obs}\sim 0$ (orange: total; green: molecular) and $1$ (cyan: total); the shaded areas illustrate the $1\sigma$ variance associated with the average over different formation redshifts. Data at $z\sim 0$ are from Peeples et al. (2014; filled squares), Saintonge et al. (2017; filled circles for total and open circles for H$_2$), Cortese et al. (2017 for starbursts; filled inversed triangles), and at $z\sim 1$ from Tacconi et al. (2013, 2018 for H$_2$; crosses with arrows); the filled and empty stars highlight the approximate locations of the Milky Way and of the Large Magellanic Cloud.}\label{fig|fgas}
\end{figure}

Rotating discs are unstable to gravitational fragmentation and conducive to molecular cloud formation, more and more efficiently as $Q$ lowers. Inspired from the above, in our one-zone framework we prescribe a linear relationship between the molecular fraction and the Toomre parameter $f_{\rm H_2} = \max[0,\omega_0-\omega_1\,Q]$, with values $\omega_0\approx 1/3$ and $\omega_1\approx 5/3$ gauged to yield a good match of the total and molecular gas to stellar mass ratios as a function of the stellar mass (see Sect.~\ref{sec|results} and Fig.~\ref{fig|fgas}). The molecular fraction is turn sets the quantity $s\equiv \tau_{\rm cond}/\tau_\star \simeq \tau_{\rm cond}\,f_{\rm H_2}/50\, t_{\rm dyn}(R_{\rm rot})$ entering our analytic solutions; in Fig.~\ref{fig|params} it is shown to take on values around unity for halo masses $M_{\rm H}\gtrsim 10^{12}\, M_\odot$ and to decrease toward smaller masses because of the drop in molecular fraction $f_{\rm H_2}$.

\subsubsection{SN feedback and wind recycling}\label{sec|wind}

The mass loading factor $\epsilon_{\rm out}$ of outflows related to type-II SNe and stellar winds can be written as (see White \& Frenk 1991; for reviews and further references see Mo et al. 2010 and Somerville \& Dave 2015)
\begin{equation}\label{eq|epsout}
\epsilon_{\rm out} = \frac{\epsilon_{\rm SN}\,\eta_{\rm SN}\, E_{\rm SN}}{E_{\rm bind}} \approx 10\,\epsilon_{SN,0.2}\, \eta_{\rm SN,-2}\, E_{\rm SN,51}\, M_{\rm H,12}^{-2/3}\, E_{z}^{-1/3}~,
\end{equation}
in terms of the occurrence $\eta_{\rm SN,-2}=\eta_{\rm SN}/0.01\, M_\odot$ of SNe per unit solar mass formed into stars, of the average energy $E_{\rm SN, 51}=E_{\rm SN}/10^{51}$ erg per single SN explosion, of the energy fraction $\epsilon_{\rm SN,0.2}=\epsilon_{\rm SN}/0.2$ effectively coupled to the interstellar medium and available to drive the outflow (e.g., MacLow \& Ferrara 1999; Murray et al. 2005; Oppenheimer \& Dave 2006; Strickland \& Heckman 2009; Heckman \& Thompson 2017; Kim \& Ostriker 2018; ), and of the gas specific binding energy $E_{\rm bind}\approx 10^{14}\, M_{\rm H,12}^{2/3}\, E_{z}^{1/3}$ cm$^2$ s$^{-2}$ in the halo potential well (see Zhao et al. 2003; Mo \& Mao 2004).

\begin{figure}
\centering
\includegraphics[height=18cm]{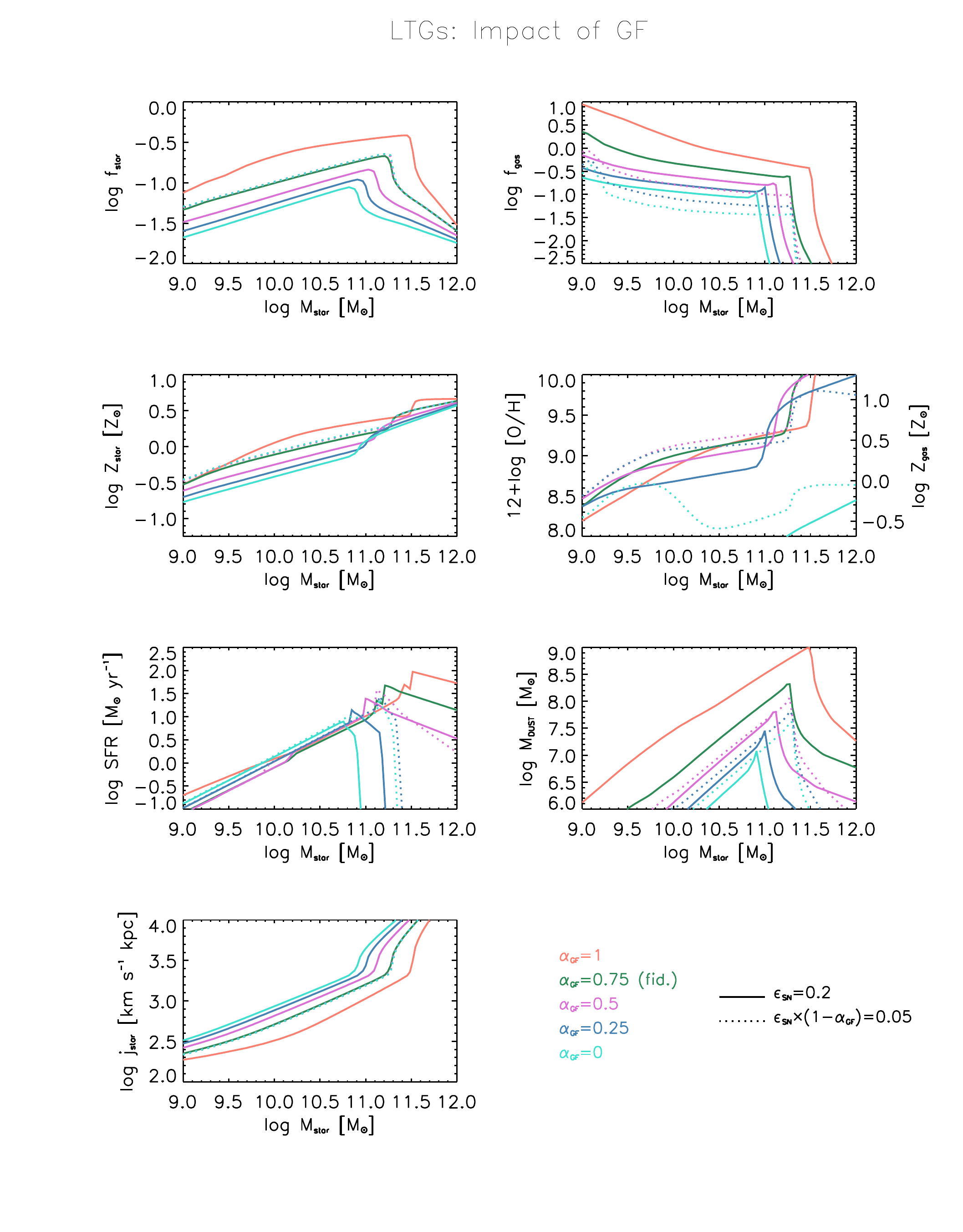}
\caption{Comparison plot showing the sensitivity of the results presented in this paper to the efficiency of the wind recycling/galactic fountain $\alpha_{\rm GF}=1$ (orange), $0.75$ (our fiducial value; green), $0.5$ (purple), $0.25$ (blue), and $0$ (cyan). Solid lines refer to our fiducial value of the stellar feedback efficiency $\epsilon_{\rm SN}=0.2$, while dotted lines (barely visible and superimposed to the solid ones in some of the plots) refer to $\epsilon_{\rm SN}\, (1-\alpha_{\, GF})=0.05$.}\label{fig|complot}
\end{figure}

We adopt a fiducial value of the wind recycling fraction $\alpha_{\rm GF}\approx 0.75$, consistent with the outcomes of several numerical simulations including wind recycling and galactic fountains (e.g., Oppenheimer et al. 2010; Ubler et al. 2014; Nelson et al. 2015; Christensen et al. 2016; Angles-Alcazar et al. 2017; Grand et al. 2019; Tollet et al 2019). In fact, some of these works predict a mass dependence of $\alpha_{\rm GF}$, which is somewhat expected since gas outflows driven by stellar feedback can more easily escape the potential wells of smaller halos with $M_{\rm H}\lesssim$ some $10^9\, M_\odot$, so reducing the retained fraction there. However, being mainly interested in halo masses larger than these values, we prefer not to introduce additional parameters describing a halo mass dependence, and to take $\alpha_{\rm GF}$ as a constant. We anticipate that the adopted value $\alpha_{\rm GF}\approx 0.75$ used in our analytic solutions will yield the best matching with the observational relationship considered in Sect.~\ref{sec|results}; there we will also discuss and illustrate in Fig.~\ref{fig|complot} how our results will be affected by variations of $\alpha_{\rm GF}$.

Note that, with respect to the standard value of the feedback efficiency $\epsilon_{\rm SN}\approx 0.05$ adopted in models without galactic fountains (see aforementioned references), here we use a higher value $\epsilon_{\rm SN}\approx 0.2$. This is reasonable, since from Eq.~(\ref{eq|fstar}) it is seen that the combination of parameters $\epsilon_{\rm out}\, (1-\alpha_{\rm GF})\propto \epsilon_{\rm SN}\, (1-\alpha_{\rm GF})$ constitutes the effective efficiency in models with fountains; given our fiducial value $\alpha_{\rm GF}\approx 0.75$, the effective efficiency $\epsilon_{\rm SN}\, (1-\alpha_{\rm GF})\approx 0.05$ assumes the standard value. The outcomes for the mass loading $\epsilon_{\rm out}$ and for the effective mass loading $\epsilon_{\rm out}\, (1-\alpha_{\rm GF})$ are illustrated in Fig.~\ref{fig|params} as a function of halo mass and formation redshift, and the latter is found to agree with self-consistent
hydrodynamical simulations of stellar feedback (e.g., Hopkins et al. 2012).

\subsubsection{IMF and metal/dust yields}\label{sec|imf}

Adopting the Chabrier (2003, 2005) IMF and the Romano et al. (2010) stellar yield models, we find an instantaneous recycling fraction $\mathcal{R}\approx 0.45$, an average yield of instantaneously produced metals $y_Z\approx 0.06$ and an yield of oxygen $y_O\approx 0.04$ (see also Krumholz et al. 2012; Nomoto et al. 2013; Feldmann 2015; Vincenzo et al. 2016). In the treatment of dust production, we adopt a dust yield $y_D\approx 7\times 10^{-4}$ (see Bianchi \& Schneider 2007; Zhukovska et al. 2008; Feldmann 2015), a strength parameter $\kappa_{\rm SN}\approx 10$ for dust spallation by SN winds (see McKee 1989; de Bennassuti et al. 2014), and an efficiency for dust accretion $\epsilon_{\rm acc}\approx 10^6$ (see Hirashita 2000; Asano et al. 2013; Feldmann 2015); these parameters are rather uncertain, and mainly set basing on previous literature and on obtaining a good match to the dust vs. stellar mass relationship at $z\sim 0-1$.

\section{Mergers and cosmic accretion}\label{sec|halogrowth}

After formation, both the DM halo and the associated baryon content are expected to grow because of mergers and smooth accretion from the cosmic web. To describe these processes, we rely on the outcomes of $N-$body simulations, and in particular the \texttt{Illustris} project (see \url{http://www.illustris-project.org/}).

The merger rates per descendant halo, per unit cosmic time and per halo mass ratio $\mu_{\rm H}$ can be described with the fitting formula originally proposed by Fakhouri \& Ma (2008), Fakhouri et al. (2010) and Lapi et al. (2013)
\begin{equation}\label{eq|halomergerate}
\frac{{\rm d}N_{\rm H, merg}}{{\rm d}t\, {\rm d}\mu_{\rm H}} = N_{\rm H}\, M_{\rm H,12}^a\, \mu_{\rm H}^{-b-2}\, e^{(\mu_{\rm H}/\tilde\mu_{\rm H})^c}\, \frac{{\rm d}\delta_c}{{\rm d}t}
\end{equation}
in terms of the descendant halo mass $M_{\rm H,12}=M_{\rm H}/10^{12}\, M_\odot$, and of the linear threshold for collapse $\delta_c(z)$.
Genel et al. (2010) have determined the parameters entering the above expression from the \texttt{Illustris}-Dark simulations, finding $N_{\rm H}=0.065$ Gyr$^{-1}$, $a=0.15$, $b=-0.3$, $c=0.5$ and $\tilde\mu_{\rm H}=0.4$. The average relative halo mass growth $\langle\dot M_{\rm H\, merg}\rangle/M_{\rm H}$ is obtained by multiplying the above expression by $\mu_{\rm H}/(1+\mu_{\rm H})$ and integrating over $\mu_{\rm H}$ from a minimum value $\mu_{\rm H, min}$ (see Rodriguez-Gomez et al. 2010); we use $\mu_{\rm H, min}\approx 10^{-5}$ that corresponds to include all mergers, since the contribution from smaller mass ratios to the halo mass growth is essentially negligible.

\begin{figure}
\centering
\includegraphics[height=18cm]{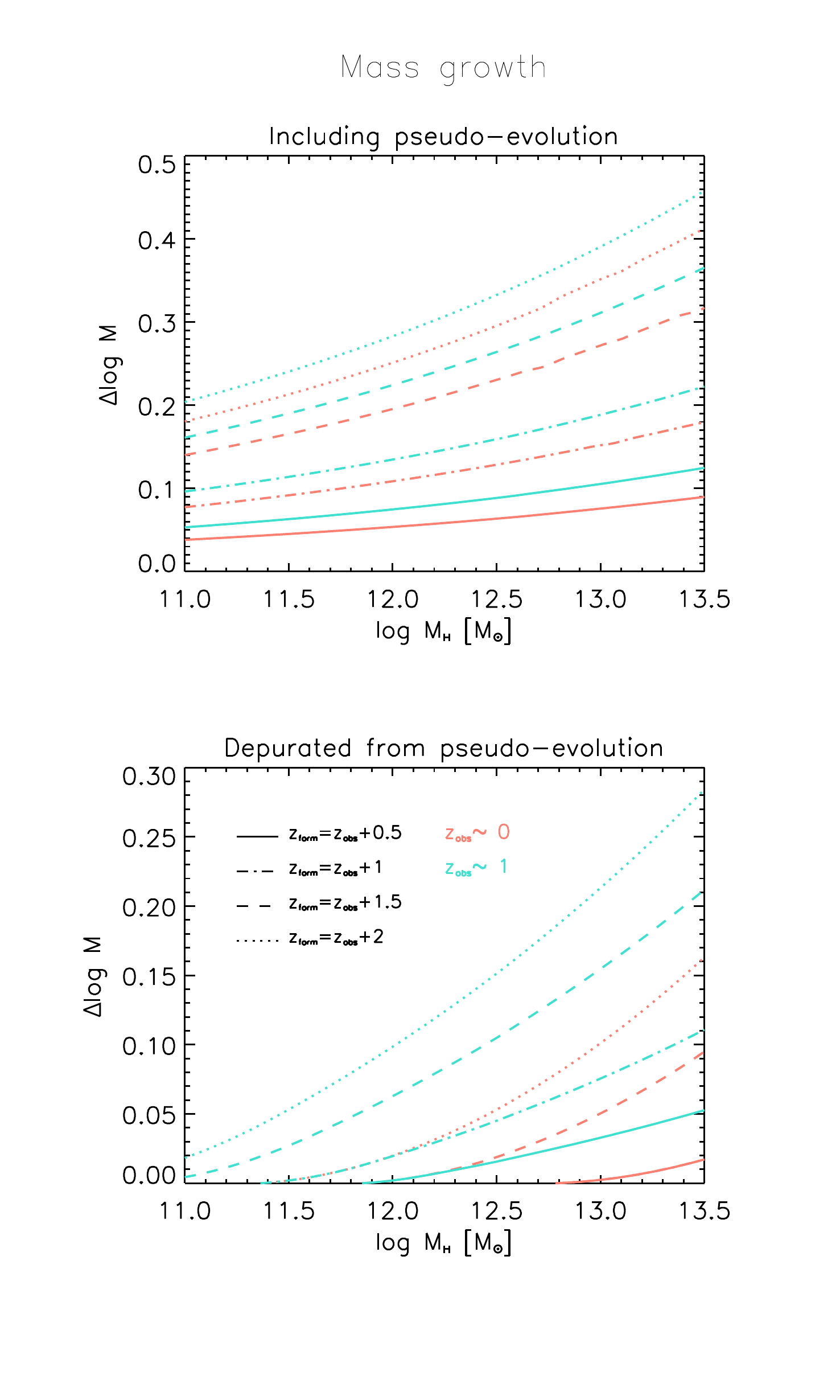}
\caption{Mass growth of DM halos (and of the baryon content) as a function of the final halo mass. Top panel refers to the mass within the standard virial overdensity radius, which is subject to pseudo-evolution due to the lowering of the background density; bottom panel refers to the mass accretion depurated from pseudo evolution (see Sect.~\ref{sec|pseudoev} for details). Results are illustrated for observation reshifts $z_{\rm obs}=0$ (orange lines) and $z_{\rm obs}=1$ (cyan lines), and for different formation redshifts $z_{\rm form}=z_{\rm obs}+0.5$ (solid lines), $z_{\rm obs}+1$ (dot-dashed lines), $z_{\rm obs}+1.5$ (dashed lines), and $z_{\rm obs}+2$ (dotted lines).}\label{fig|halogrowth}
\end{figure}

During each timestep ${\rm d}t$ after halo formation at $z_{\rm form}$ till the observation redshift $z_{\rm obs}$, the halo mass is increased by the amount $\langle \dot M_{\rm H,merg}\rangle\, {\rm d}t$ (and the associated baryonic mass by the amount $f_{\rm b}\,\langle \dot M_{\rm merg,H}\rangle\, {\rm d}t$), to obtain the overall evolution of the halo $M_{\rm H}(z)$ and baryonic mass $M_{\rm b}(z)$; since the merger rates are small (and even more so when depurated from pseudo-evolution, see below), Eqs.~(4) can be considered still valid quasi steady-state solutions, with a time-dependent $M_{\rm b}$. To have a more quantitative grasp on such mass additions, in Fig.~\ref{fig|halogrowth} (top panel) we illustrate the growth of the halo mass as a function of the descendant final masses. We show the outcomes at observation redshifts $z_{\rm obs} = 0$ and $1$ (color coded), and different formation redshifts $z_{\rm form}=z_{\rm obs}+0.5$, $z_{\rm obs}+1$, $z_{\rm obs}+1.5$, $z_{\rm obs}+2$ (linestyle coded). Plainly, at given observation and formation redshift the amount of relative mass additions increases with the descendant mass, while at given descendant mass and observation redshift the mass additions increase for higher formation redshift; the dependence on the observation redshift is weak.

\subsubsection{Pseudo-evolution}\label{sec|pseudoev}

However, it has been pointed out by several authors (see Diemand et al. 2005; Cuesta et al. 2008; Zemp 2014; More et al. 2015; Diemer et al. 2013, 2017) that after formation the halo growth is mainly driven by a pseudo-evolution in radius and mass due to the lowering of the reference density defining the halo boundary. For example, pseudo-evolution implies that a halo of $10^{13}\, M_\odot$ at a formation redshift $z_{\rm form}\approx 2$ will end up, say, at $z\approx 0$ in a halo of several $10^{14}\, M_\odot$; clearly the latter is a halo typical of a galaxy group/cluster, but has little relevance to the physical processes occurring within the galaxy hosted by the original halo of $10^{13}\, M_\odot$ at formation. In other words, the `galactic' halo must have evolved in mass much less than predicted by pseudo-evolution. Under the assumption that the universal NFW mass distribution $M(<r)\propto \ln(1+c\,x)-c\,x/(1+c\,x)$ with $x\equiv r/R_{\rm H}$ is retained, the mass growth $M_{\rm GH}(z)$ of the galactic halo (depurated from pseudo-evolution) can be estimated as (see More et al. 2015; Diemer et al. 2017)
\begin{equation}\label{eq|galhaloevo}
M_{\rm GH}(z)\simeq M_{\rm H}(z)\,  \cfrac{\ln(1+c_{\rm form})-c_{\rm form}/(1+c_{\rm form})}{\ln(1+c)-c/(1+c)}
\end{equation}
where $c\approx c_{\rm form}\, (1+{z_{\rm form}})/(1+z)$ and $c_{\rm form}\approx 4$; other prescriptions based, e.g., on the virial ratio yield similar behaviors (e.g., Zemp 2014). In Fig.~\ref{fig|halogrowth} (bottom panel, note the change in y-axis scale with respect to top panel) we illustrate the growth of the galactic halo mass as a function of the descendant final masses. The halo growth depurated from pseudo-evolution is weaker than the standard one for a given observation redshift, and especially so for low mass halos.

For example, the halo mass increase at $z_{\rm obs}\approx 0$ for a descendant halo mass $M_{\rm H}\approx 10^{12}\, M_\odot$ ($10^{13}\, M_\odot$) formed at $z_{\rm form}\approx 2$ amounts to a factor $\approx  2$ ($\approx  3$) when including pseudo-evolution, but reduces to less than $15\%$ ($30\%$) when depurated from it; conversely, a halo mass of $10^{12}\, M_\odot$ ($10^{13}\, M_\odot$) at $z_{\rm form}\approx 2$ will have increased its mass at $z_{\rm obs}\approx 0$ by a factor $\approx 3$ ($\approx 4$) when including pseudo-evolution, but only of $\approx 40\%$ ($\approx 2$) when depurating from it. Thus essentially galactic halos hosting LTGs are subject to little evolution after formation; this is particularly relevant for the stability of these systems, since discs are known to constitute fragile structures, that would be easily destroyed by a rain of substantial merging events after formation.

\subsubsection{Average over formation redshifts}\label{sec|average}

In order to derive the statistical properties of the galaxy population concerning a quantity $\mathcal{Q}$ (e.g., star formation rate, stellar mass), we proceed as follows. We exploit the analytic solutions of Sect.~\ref{sec|ansol} for the evolution of individual galaxies with different formation redshift $z_{\rm form}$ and halo masses $M_{\rm H}$ at formation, to obtain $\mathcal{Q}(\tau|M_{\rm H},z_{\rm form})$. Given an observation redshift $z_{\rm obs}\la z_{\rm form}$, we pick up the value of $\mathcal{Q}$ at $\tau=t_{z_{\rm obs}}-t_{z_{\rm form}}$, where $t_z$ is the cosmic time at redshift $z$. Finally, we perform the average over different formation redshift to get
\begin{equation}\label{eq|zformave}
\langle \mathcal{Q}\rangle(M_{\rm H},z_{\rm obs})\propto \int^{\infty}_{z_{\rm obs}}{\rm d}z_{\rm form}\, \frac{{\rm d}^2N_{\rm H}}{{\rm d}\log M_{\rm H}\, {\rm d}z_{\rm form}}\, \mathcal{Q}(t_{z_{\rm obs}}-t_{z_{\rm form}}|M_{\rm H},z_{\rm form})
\end{equation}
where ${\rm d}^2N_{\rm H}/{\rm d}\log M_{\rm H}\, {\rm d}z_{\rm form}$ is the halo formation rate computed via the excursion set framework, and gauged against $N-$body simulations, by Lapi et al. (2013; see also Lacey \& Cole 1993; Kitayama \& Suto 1996; Moreno et al. 2009; Giocoli et al. 2012). The normalization constant in the above Eq.~(\ref{eq|zformave}) is clearly the same integral without $\mathcal{Q}$, and the $1\sigma$ variance is computed as $\sigma_{\mathcal{Q}}=\sqrt{\langle \mathcal{Q}^2\rangle-\langle \mathcal{Q}\rangle^2}$.

To compute the formation rate one needs the halo mass function, i.e., the co-moving number density of halo per halo mass bin, as provided by state-of-the-art N-body simulations; we adopt the determination by Tinker et al. (2008; see also Watson et al. 2013; Bocquet et al. 2016; Comparat et al. 2017, 2019). Since we are mainly concerned with the properties of galactic halos hosting LTGs, we actually exploit the galactic halo mass function, i.e., the mass function of halos hosting one individual galaxy (though the difference with respect to the halo mass function emerge only for $z\la 1$ and $M_{\rm H}\ga$ several $10^{12}\, M_\odot$). This can be built up from the overall halo mass function by adding to it the contribution of sub-halos and by probabilistically removing from it the contribution of halos corresponding to galaxy systems via halo occupation distribution modeling; we refer the reader to Appendix A of Aversa et al. (2015) for details on such a procedure.

\section{Results}\label{sec|results}

\begin{figure}
\centering
\includegraphics[width=\textwidth]{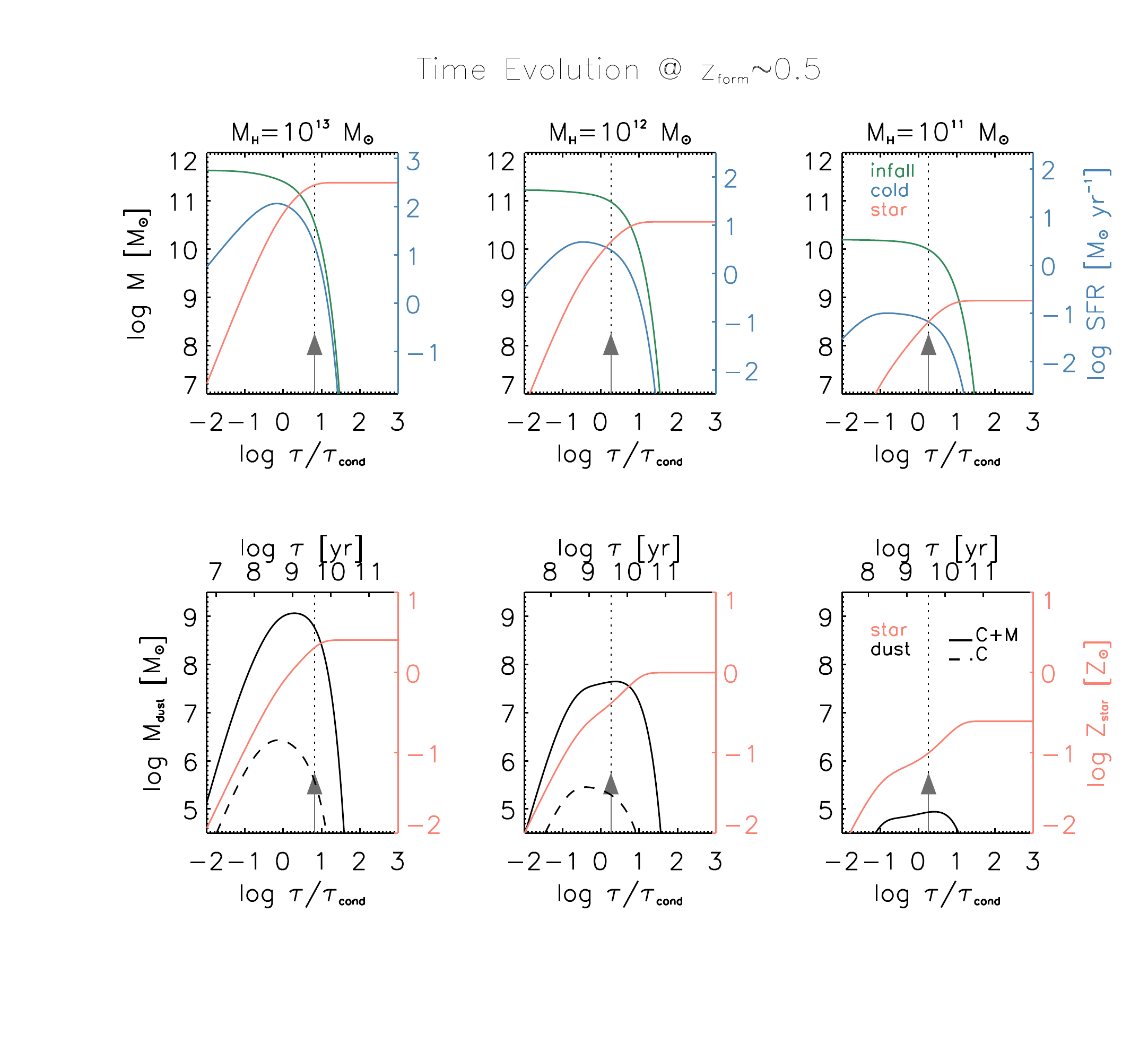}
\caption{Top panels: evolution of the infall (green), cold (blue), and stellar (orange) masses as a function of galactic age $\tau$ (normalized to the condensation timescale $\tau_{\rm cond}$ or in absolute units of yr) for galaxies hosted in halos with mass $M_{\rm H}= 10^{13}\, M_\odot$ (left panels), $M_{\rm H}= 10^{12}\, M_\odot$ (middle panels), and $M_{\rm H}= 10^{11}\, M_\odot$ (right panels) at a
formation redshift $z_{\rm form}=0.5$; the right $y-$axis reports the SFR (in log-scale) and refers to the cyan line. Bottom panels: evolution of the dust mass (black, refers to the left $y-$axis; dashed line is for the core and solid for core plus mantles) and of the stellar metallicity (orange, refers to the right $y-$axis). In all panels, the vertical arrow and dotted line indicates the present time $t(z=0)-t(z_{\rm form})$.}\label{fig|timevo_z05}
\end{figure}

\begin{figure}
\centering
\includegraphics[width=\textwidth]{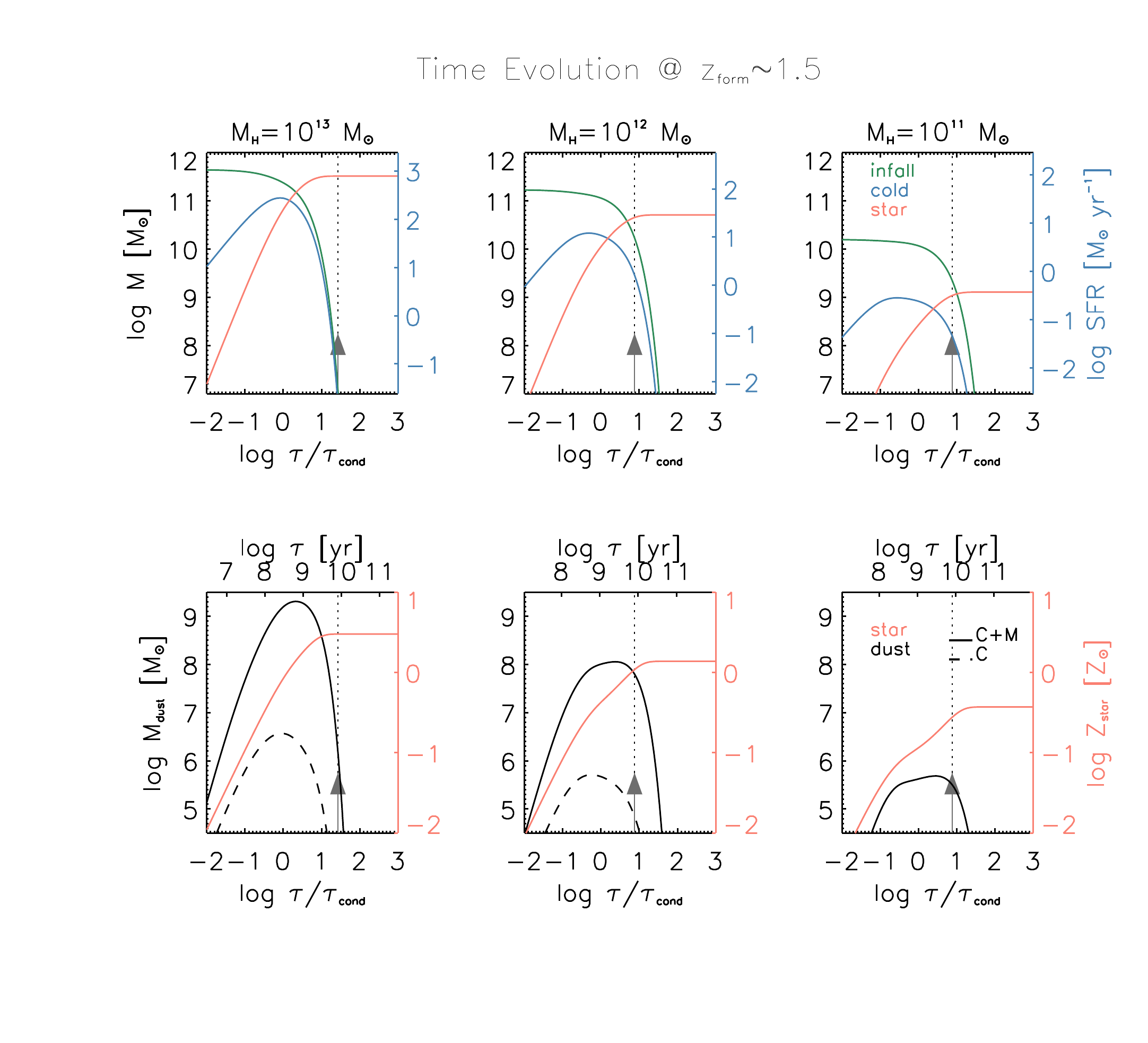}
\caption{Same as previous figure for a formation redshift $z_{\rm form}\sim 1.5$.}\label{fig|timevo_z15}
\end{figure}

In Figs.~\ref{fig|timevo_z05} and \ref{fig|timevo_z15} we illustrate the evolution with galactic age $\tau$ of the relevant spatially-averaged quantities described by our analytic solutions: infalling gas mass, cold gas mass, stellar mass, dust mass and stellar metallicity. Each figure refers to a different formation redshift $z_{\rm form}\approx 0.5$ and $1.5$ typical for LTGs, and depicts the evolution for representative galaxies with halo masses $M_{\rm H}=10^{11}-10^{12}-10^{13}\,M_\odot$ at formation.
We have included external halo and baryonic accretions, though when depurated from pseudo-evolution their contribution to the quantities illustrated in Figs.~\ref{fig|timevo_z05}-\ref{fig|timevo_z15} is minor, especially toward higher $z_{\rm form}$.

The infalling gas mass decreases almost exponentially with the galactic age as it condenses in the cold gas component, though it is partially refurnished by wind recycling; the cold gas then feeds star formation and is affected by stellar feedback. The balance of these processes makes the cold gas mass (hence the SFR) to slowly grow at early times, to attain a maximum and then to decrease almost exponentially; correspondingly, the stellar component increases almost linearly and then saturates. In less massive halos stellar feedback is more efficient, the initial gas reservoir is smaller, and the molecular fraction is lower (hence the star formation timescales are longer); at the same time, wind recycling is more effective in refurnishing the infall gas reservoir and thus extending the duration of the star formation. The stellar metallicity rises almost linearly at early times, then slows down because of the balance between dilution, feedback and the recycling of enriched gas by the fountain, and then saturates to an asymptotic value reflecting the effective stellar yield. Moving toward lower mass halos the saturation values are progressively smaller since more metals are ejected by the action of stellar feedback. The dust mass features a behavior similar to the cold gas phase, with an extended maximum attained as soon as metals are available for accretion onto the grain cores, and then decreases as the gas mass gets exhausted. Looking at  Figs.~\ref{fig|timevo_z05}-\ref{fig|timevo_z15} it is seen that for a typical Milky-Way sized galaxy with halo mass $M_{\rm H}\approx 10^{12}\, M_\odot$ formed at $z\lesssim 1.5$ we expect a present day stellar mass of $M_\star\approx$ some $10^{10}\, M_\odot$, a current SFR $\approx $a few $M_\odot$ yr$^{-1}$ and a stellar metallicity $Z_\star\approx Z_\odot$. For smaller galaxies like the Large Magellanic Cloud (LMC) with halo masses $M_{\rm H}\approx 10^{11}\, M_\odot$ formed at $z\gtrsim 1.5$ we find $M_\star \lesssim 10^{9}\, M_\odot$, SFR $\lesssim 10^{-1}\,M_\odot$ yr$^{-1}$ and $Z_\star\approx 0.3\, Z_\odot$. Notice that very massive halos with $M_{\rm H}\approx 10^{13}\, M_\odot$ and $z_{\rm form}\gtrsim 1.5$ feature present-day stellar masses $M_\star\gtrsim 10^{11}\, M_\odot$ but tend to be caught in the exponentially declining phase of their star formation histories; in our framework, such very large halos rarely host a local star-forming LTGs (as mentioned below, this concurs with indications from weak lensing measurements), and can do so only if formed relatively recently at $z_{\rm form}\lesssim 1$.

\begin{figure}
\centering
\includegraphics[width=\textwidth]{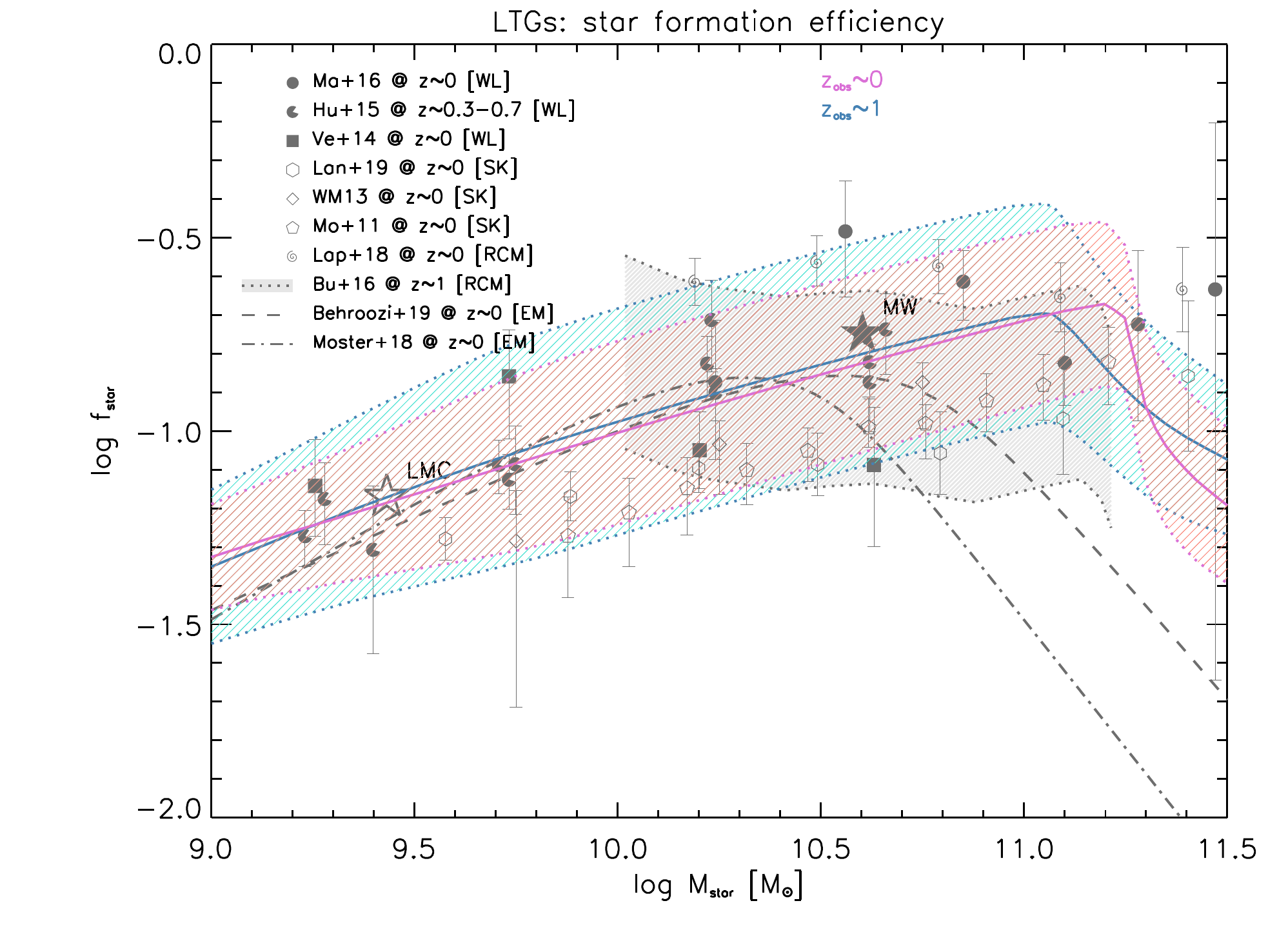}
\caption{Star formation efficiency $f_\star\equiv M_\star/f_{\rm b}\, M_{\rm H}$ as a function of the stellar mass $M_\star$, at observation redshifts $z_{\rm obs}\sim 0$ (orange) and $1$ (cyan); the shaded areas illustrate the $1\sigma$ variance associated with the average over different formation redshifts. Data are from Mandelbaum et al. (2016; circles), Hudson et al. (2015; pacmans), Velander et al. (2014; squares) via weak lensing measurements; Lange et al. (2019; open hexagons), Wojtak \& Mamon (2013; open diamonds), and More et al. (2011; open pentagons) via satellite kinematics; Lapi et al. (2018b; spirals) at $z\sim 0$ and Burkert et al. (2016; grey shaded area) at $z\sim 1$ via rotation curve modeling; the filled and empty stars highlight the approximate locations of the Milky Way and of the Large Magellanic Cloud. The results from the empirical models by Moster et al. (2018; dot-dashed line) and Behroozi et al. (2019; dashed line) are also reported.}\label{fig|fstar}
\end{figure}

In Fig.~\ref{fig|fstar} we illustrate the star formation efficiency $f_{\star}\equiv M_{\star}/f_{\rm b}\, M_{\rm H}$, namely the fraction of baryonic mass converted into stars, as a function of the stellar mass for two different observation redshifts $z_{\rm obs}=0$ and $1$ relevant for LTGs (color-coded); the shaded areas illustrate the 1$\sigma$ scatter associated to the average over different formation redshifts. The star formation efficiency $f_{\star}$ is a non-monotonic function of the stellar mass $M_{\star}$, with a maximum value of $20\%$ around $M_{\star}\simeq 10^{11}\, M_\odot$, and a decrease to values less than $10\%$ for $M_{\star}\lesssim$ some $10^9\, M_\odot$ and for $M_{\star}\gtrsim$ a few $10^{11}\, M_\odot$. These values basically imply that star formation in LTGs is a very inefficient process.
Such a behavior as a function of the stellar mass is easily understood in terms of infall/condensation and feedback processes: toward small masses, star formation is more easily offset by stellar feedback, which ejects gas more easily from the shallower potential wells of these systems; conversely, toward larger masses the infall fraction become progressively smaller since the cooling times get longer. In analytic terms,
from Eq.~(\ref{eq|epsout}) the feedback mass loading factor is seen to scale as $\epsilon_{\rm out}\propto M_{\rm H}^{-2/3}$, and after Eq.~(\ref{eq|fstar}) this implies $f_\star\propto f_{\rm inf}$ at high masses and $f_\star\propto f_{\rm inf}^{3/5}\,(1-\alpha_{\rm GF})^{-3/5}\,M_{\star}^{2/5}$ at small masses. Our results are compared with the data for LTGs from various works, relying on weak lensing (see Velander et al. 2014; Hudson et al. 2015; Mandelbaum et al. 2016), satellite kinematics (see More et al. 2011; Wojtak \& Mamon 2013; Lange et al. 2019), rotation curve modeling (Lapi et al. 2018 at $z\approx 0$ and Burkert et al. 2016 at $z\approx 1$), and empirical modeling based on abundance matching and halo occupation distribution (see Moster et al. 2018; Behroozi et al. 2019). We find a very good agreement in normalization and scatter with these determinations, within the still large observational uncertainties.

\begin{figure}
\centering
\includegraphics[width=0.7\textwidth]{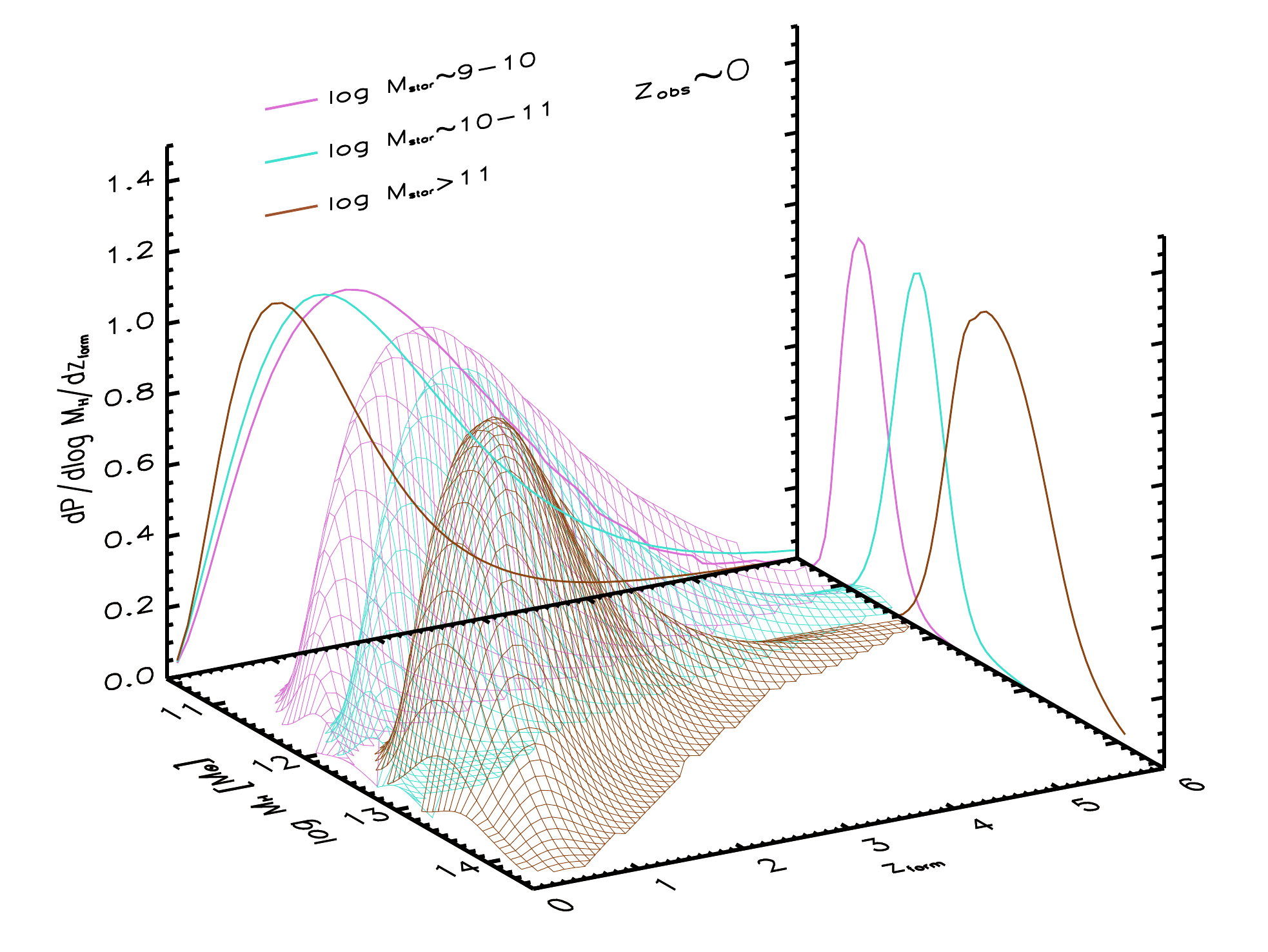}
\includegraphics[width=0.7\textwidth]{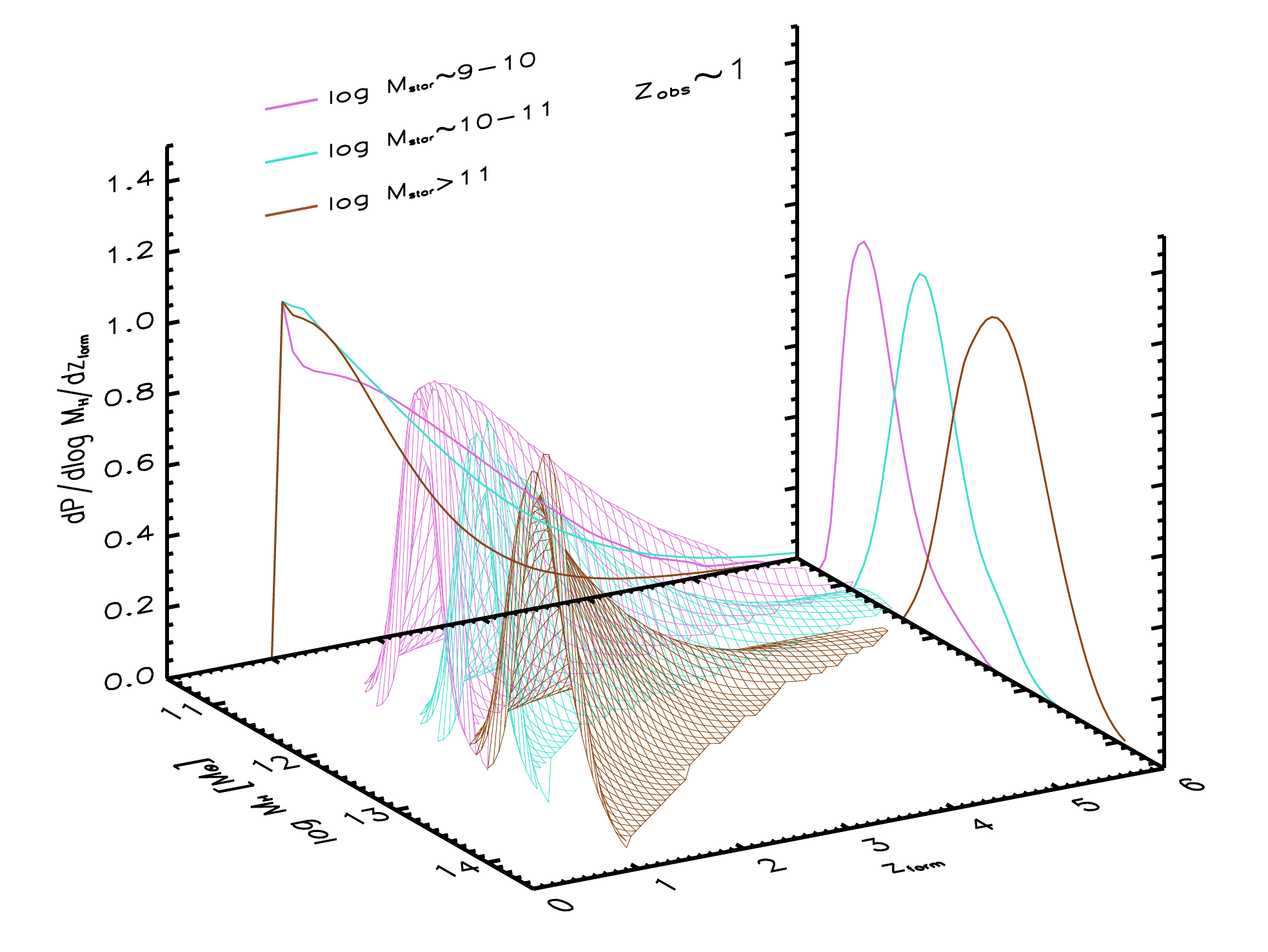}
\caption{Joint and marginalized distributions of halo masses and formation redshifts in different stellar mass bins $\log M_\star/M_\odot\approx 9-10$ (magenta), $10-11$ (cyan), and $>11$ (brown) at observation redshift $z_{\rm obs}\approx 0$ (top panel) and $\approx 1$ (bottom panel).}\label{fig|prob3D}
\end{figure}

In Fig.~\ref{fig|prob3D} we show the joint and marginalized distributions of halo mass and formation redshift for galaxies in three bins of stellar masses $M_\star\sim 10^{9-10}\, M_\odot$, $\sim 10^{10-11}\, M_\odot$, and $\gtrsim 10^{11}$ (color-code) at two different observation redshifts $z_{\rm obs}\approx 0$ (top panel) and $\approx 1$ (bottom panel). Focusing on local galaxies at $z_{\rm obs}\approx 0$, the distribution of halo masses is roughly lognormal, with a dispersion around $0.2-0.3$ dex and a median corresponding to the star-formation efficiency discussed above and illustrated in the previous Figure. As for the distribution in formation redshift, it roughly peaks in the range $z_{\rm form}\approx 1-2$, and features a tail toward higher $z_{\rm form}$, more extended for galaxies with smaller stellar masses. Note that recurrent bursts of star formation toward the present (not included in our framework), especially in small mass galaxies, could make the observed luminosity-averaged age of these systems appreciably shorter than what can be inferred from the intrinsic formation time distribution presented here. At $z_{\rm obs}\approx 1$ the distributions in halo masses are similar, while those in formation redshift peak around $z_{\rm form}\lesssim 1.2$ and decline quite steeply for higher $z_{\rm form}$; the latter behavior is enforced by a combination of the shape in the halo formation rates, and of the time evolution in the stellar mass growth for individual galaxies observed at $z_{\rm obs}\approx 1$. Note that the small kink toward $z_{\rm form}\approx z_{\rm obs}\approx 1$ in the redshift distribution (particularly evident for the lower stellar mass bin) is caused by young objects which, being hosted in large halo masses, are rapidly crossing the stellar mass bin but will end up with much higher relic stellar masses at the present time.

In Fig.~\ref{fig|fgas} we illustrate the relationship between the gas to star mass fraction $f_{\rm gas}\equiv M_{\rm gas}/M_{\star}$ and the stellar mass $M_{\star}$ for two different observation redshifts $z_{\rm obs}\sim 0$ and $1$ relevant for LTGs (color-coded). The gas mass fraction is found to increase monotonically toward smaller stellar masses. Such a behavior is essentially driven by the decrease of the molecular gas fraction $f_{\rm H_2}$ toward smaller halo masses (see below Eq.~\ref{eq|Toomre}); in fact, Eq.~(\ref{eq|latebasicsol}) implies $f_{\rm gas}\propto s^{-1}\, F(\lambda_\pm)\propto f_{\rm H_2}^{-1}$ since $s\propto f_{\rm H_2}$ applies and the residual factor $F(\lambda_\pm)$ depends very weakly on mass. Our results on the total and molecular gas mass fractions are compared with the observational determinations by Peeples et al. (2014), Cortese et al. (2017; for starbursts), Saintonge et al. (2017; see also Catinella et al. 2018 and Calette et al. 2018) at $z\sim 0$, and by Tacconi et al. (2013, 2018; only H$_2$) at $z\sim 1$. A broad agreement with the data within their uncertainties and intrinsic variance is found.

\begin{figure}
\centering
\includegraphics[width=\textwidth]{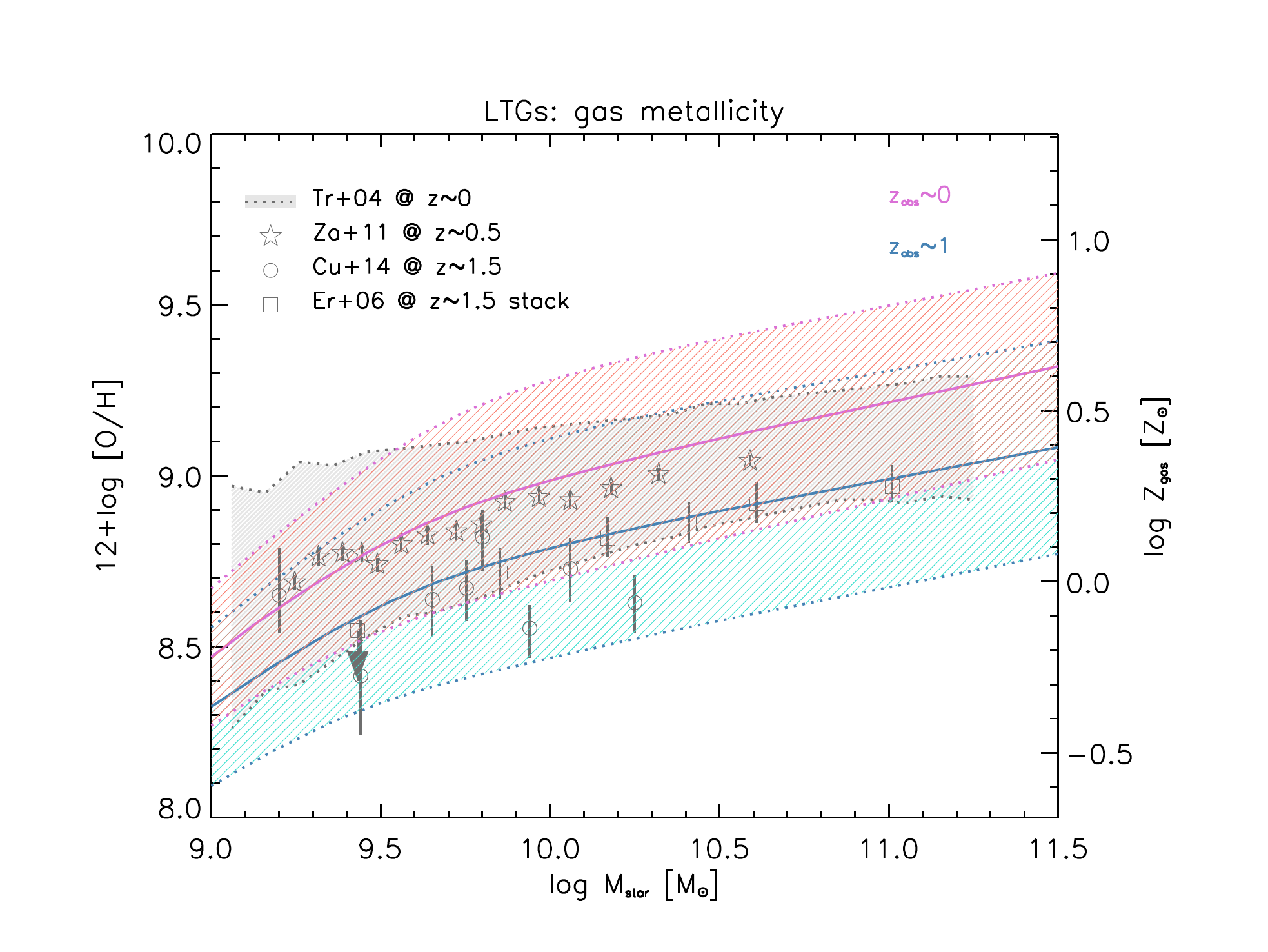}
\caption{Gas metallicity $12+\log[O/H]$ (left $y$-axis) or equivalently $Z_{\rm gas}$ (right $y$-axis) as a function of the stellar mass $M_\star$, at observation redshifts $z_{\rm obs}\sim 0$ (orange) and $1$ (cyan); the shaded areas illustrate the $1\sigma$ variance associated with the average over different formation redshifts. Data are from Tremonti et al. (2004; grey shaded area) at $z\sim 0$, Zahid et al. (2011; stars) at $z\sim 0.5$, Cullen et (2014; circles) and Erb et al. (2006; squares) at $z\sim 1.5$.}\label{fig|Zgas}
\end{figure}

In Fig.~\ref{fig|Zgas} we illustrate the mass-metallicity relationship (see Tremonti et al. 2004; Erb et al. 2006; Zahid et al. 2011; Cullen et al. 2014), i.e., the relation between the gas metallicity $Z_{\rm gas}$ and the stellar mass $M_{\star}$ at two different observation redshifts $z_{\rm obs}\sim 0$ and $1$ relevant for LTGs (color-coded).  The gas metallicity shows an increasing behavior as a function of the final stellar mass, related to the more efficient production of metals in galaxies with higher SFRs, that will also yield larger stellar masses. Our results are in reasonable agreement with gas metallicity estimates (traced mainly by Oxygen abundance, and converted to PP04O3N2 calibration, see Kewley \& Ellison 2008).

\begin{figure}
\centering
\includegraphics[width=\textwidth]{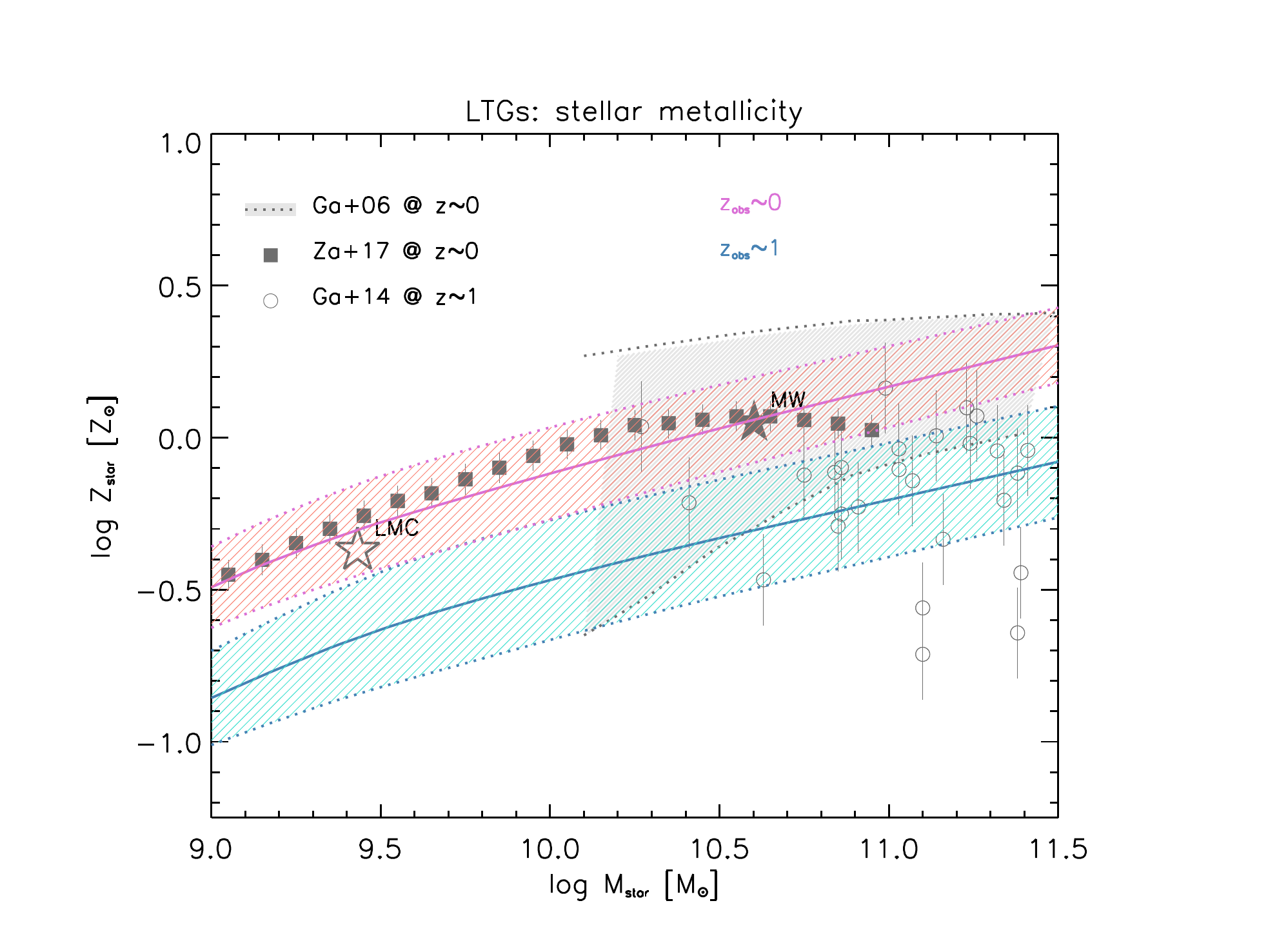}
\caption{Stellar metallicity $Z_\star$ as a function of the stellar mass $M_\star$, at observation redshifts $z_{\rm obs}\sim 0$ (orange) and $1$ (cyan); the shaded areas illustrate the $1\sigma$ variance associated with the average over different formation redshifts. Data are from Gallazzi et al. (2006; grey shaded area) and Zahid et al. (2017; stars) at $z\sim 0$, and Gallazzi et al. (2014; open circles) at $z\sim 1$; the filled and empty stars highlight the approximate locations of the Milky Way and of the Large Magellanic Cloud.}\label{fig|Zstar}
\end{figure}

In Fig.~\ref{fig|Zstar} we illustrate the stellar mass-metallicity relationship (see Gallazzi et al. 2006, 2014; Zahid et al. 2017), i.e., the relationship between the stellar metallicity $Z_{\star}$ and the stellar mass, $M_{\star}$ at two different observation redshifts $z_{\rm obs}=0$ and  $1$ relevant for LTGs (color-coded). The stellar metallicity increases monotonically with stellar mass, somewhat mirroring the gas metallicity behaviour. More massive galaxies are characterized on average by higher SFRs, that imply larger stellar masses and metal production. Moreover, in low mass galaxies the depletion of metals by stellar feedback is enhanced due to the shallower potential wells associated to the host halos. Contrariwise, high-mass galaxies can retain greater amounts of chemical-enriched gas, that could be converted and locked into new metal-rich stars, resulting in a higher stellar metallicity. Such a behavior can be easily traced back to Eq.~(\ref{eq|finf}), which implies $Z_\star\propto y_Z\, f_\star/f_{\rm inf}$ hence a modest dependence on stellar mass (see above). Our results are in good agreement with measurements of stellar metallicity in local LTGs by Gallazzi et al. (2006) and Zahid et al. (2017), and with the estimates by Gallazzi et al. (2014) at $z\sim 0.7$.

\begin{figure}
\centering
\includegraphics[width=\textwidth]{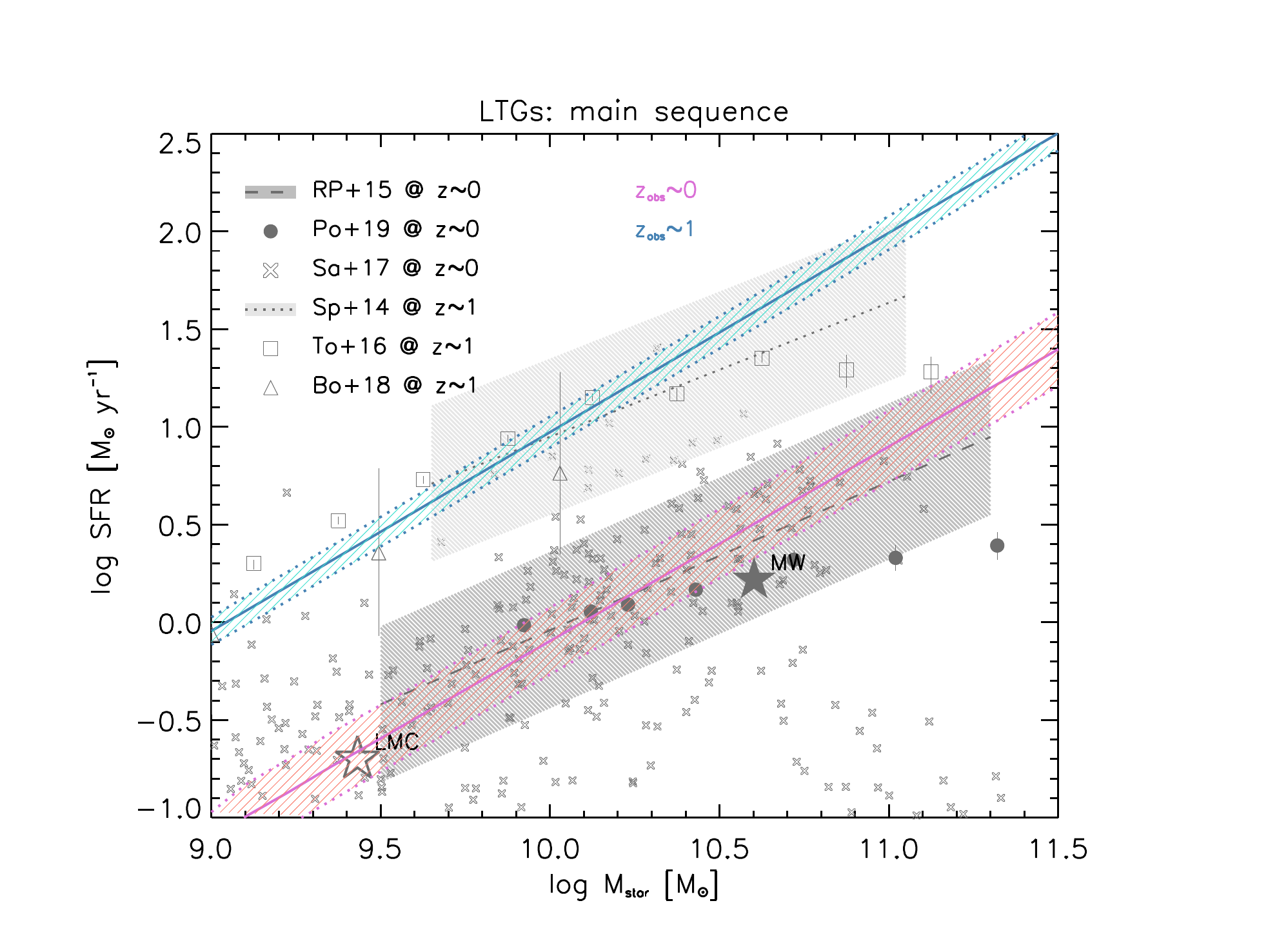}
\caption{Main sequence relationship between SFR and stellar mass $M_\star$, at observation redshifts $z_{\rm obs}\sim 0$ (orange) and $1$ (cyan); the shaded areas illustrate the $1\sigma$ variance associated with the average over different formation redshifts. Data are from Renzini \& Peng (2015; dashed line and dark shaded area), Popesso et al. (2019; filled circles), Saintonge et al. (2017; individual galaxies) at $z\sim 0$, and Speagle et al. (2014; dotted line and light shaded area),  Tomczak et al. (2016; open squares), Boogard et al. (2018, open triangles) at $z\sim 1$; the filled and empty stars highlight the approximate locations of the Milky Way and of the Large Magellanic Cloud.}\label{fig|MS}
\end{figure}

In Fig.~\ref{fig|MS} we illustrate our results concerning the so-called {\itshape main-sequence} (MS) of starforming galaxies, namely the relation between the SFR and stellar mass at two different observation redshifts $z_{\rm obs}= 0$ and $1$ relevant for LTGs (color-coded). These are confronted with the observational determinations at $z\approx 0$ by Renzini \& Peng (2015), Saintonge et al. (2017), Popesso et al. (2019) and at $z\approx 1$ by  Tomczak et al. (2016) and Boogard et al. (2018), finding a very good agreement within the large intrinsic scatter and observational uncertainties. The approximate linear shape of the relation can be easily understood from Eqs.~(\ref{eq|latebasicsol}) and (\ref{eq|SFRmax}), which imply $\dot M_\star \propto M_\star\, F(\lambda_\pm)$ in terms of a quantity $F(\lambda_\pm)$ very weakly varying with mass.
We note that the flat behavior at high stellar masses in the observational data can be partly ascribed to the inclusion of quenched systems or of starforming galaxies with a prominent bulge component, while our predictions refer to disc-dominated galaxies with ongoing star formation.

\begin{figure}
\centering
\includegraphics[width=\textwidth]{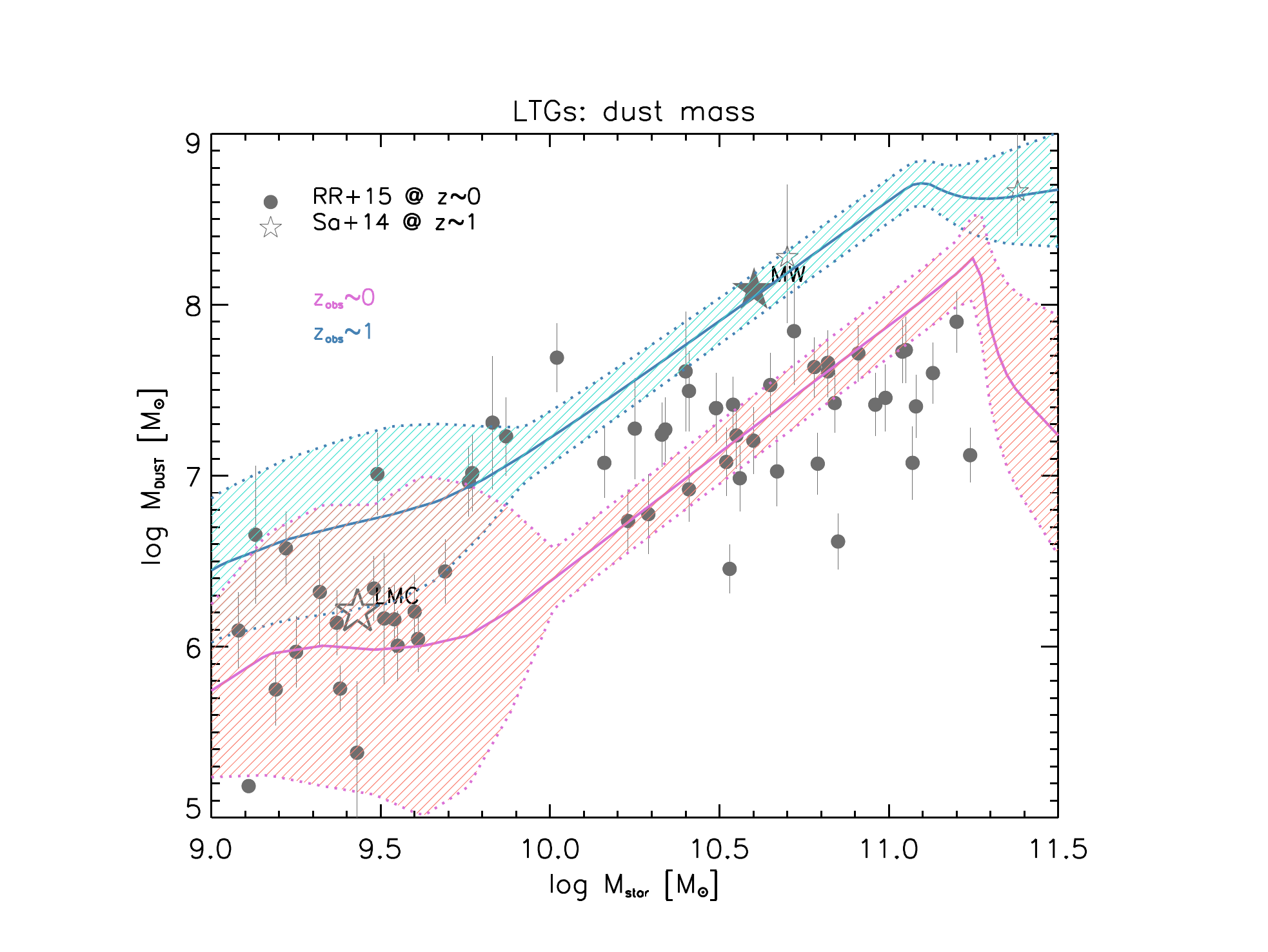}
\caption{Dust mass $M_{\rm dust}$ as a function of the stellar mass $M_\star$, at observation redshifts $z_{\rm obs}\sim 0$ (orange) and $1$ (cyan); the shaded areas illustrate the $1\sigma$ variance associated with the average over different formation redshifts. Data are from Remy-Ruyer et al. 2015 (filled circles) at $z\sim 0$, and by Santini et al. (2014; empty stars) at $z\sim 1$; the filled and empty stars highlight the approximate locations of the Milky Way and of the Large Magellanic Cloud.}\label{fig|mdust}
\end{figure}

In Fig.~\ref{fig|mdust} we illustrate the dust mass $M_{\rm dust}$ as a function of the stellar mass $M_{\star}$ for two different observation redshifts $z_{\rm obs}\sim 0$ and $1$ relevant for LTGs (color-coded). A direct relationship is expected since both dust and stellar mass are strictly related to the SFR. In fact, the dust mass is produced in stellar ejecta/winds, that are more efficient when the SFR is higher; the latter also favors metal-rich environments, in turn triggering efficient growth of dust grains by accretion. At given stellar mass, higher redshift galaxies are expected to produce slightly larger amount of dust; the trend can be traced back to their denser environment, in turn yielding shorter star formation timescales. The dust vs. stellar mass relation is easily understood on the basis of the aforementioned scaling laws, since after Eq.~(\ref{eq|Dmantleave}) the dust-to-gas ratio is seen to converge to the gas metallicity and hence the dust mass just reads $M_{\rm dust}\propto f_{\rm gas}\,Z_{\rm gas}\, M_\star\propto f_{\rm H_2}^{-1}\, Z_{\star}\, M_\star$. Our results are in good agreement with the observational estimates by Remy-Rueyer et al. (2015) at $z\sim 0$, and by Santini et al. (2014) at $z\sim 1$, though the uncertainties are still large.

\begin{figure}
\centering
\includegraphics[width=\textwidth]{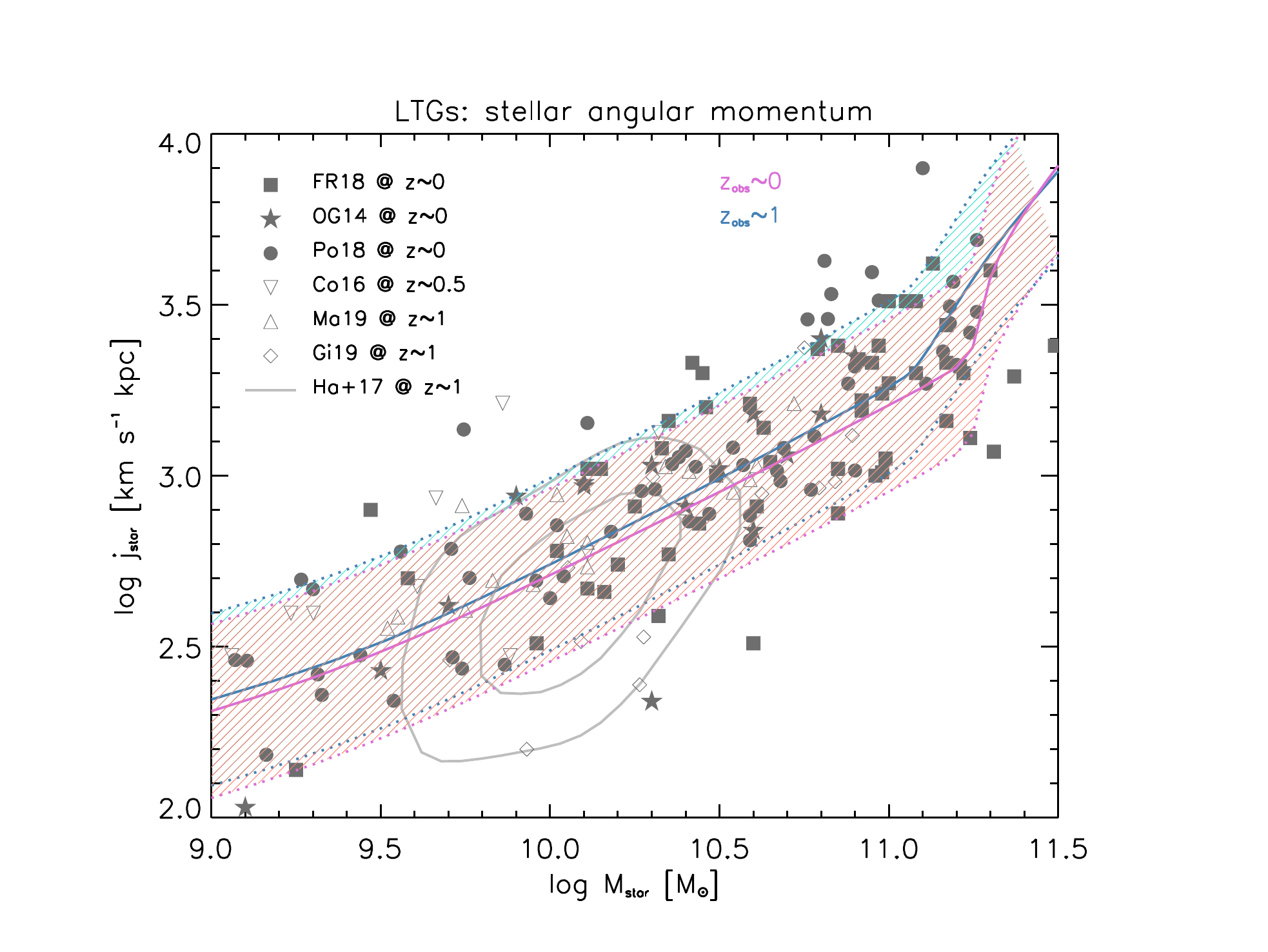}
\caption{Stellar specific angular momentum $j_\star$ as a function of the stellar mass $M_\star$, at observation redshifts $z_{\rm obs}\sim 0$ (orange) and $1$ (cyan); the shaded areas illustrate the $1\sigma$ variance associated with the average over different formation redshifts. Data are from Fall \& Romanowsky (2013; filled squares), Obreschkow \& Glazebrook (2014; filled stars) and Posti et al. (2018; filled circles) at $z\sim 0$, and from Contini et al. (2016; open reversed triangles), Marasco et al. (2019; open triangles), Gillmann et al. (2019; open diamonds), and Harrison et al. (2017; grey contours) at $z\sim 1$.}\label{fig|jstar}
\end{figure}

In Fig.~\ref{fig|jstar} we illustrate the specific angular momentum $j_\star$ in the stellar component as a function of the stellar mass $M_{\star}$ for two different observation redshifts $z_{\rm obs}\sim 0$ and $1$ relevant for LTGs (color-coded). Our expectations are in very good agreement in normalization, slope and redshift (in-)dependence with the data from Fall \& Romanowsky (2013), Obreschkow \& Glazebrook (2014) and Posti et
al. (2018) at $z\sim 0$, and from Contini et al. (2016),
Marasco et al. (2019), Gillmann et al. (2019), and Harrison
et al. (2017) at $z\sim 1$. Such a relationship basically reflects the original dependence of the halo specific angular momentum $j_{\rm H}\propto M_{\rm H}^{2/3}$ with the halo mass after Eq.~(\ref{eq|jhalo}). In LTGs, thanks to the baryon cycle enforced by cooling, stellar feedback and galactic fountains, the gas and hence the stellar mass sample well the angular momentum of the baryons originally present in the halo (Romanowsky \& Fall 2012; Fall \& Romanowsky 2018). In such a case using Eq.~(\ref{eq|fstar}) one can write (see also Shi et al. 2017)
$j_\star \propto f_{\rm inf}\, f_\star^{-2/3}\, M_\star^{2/3}$;
since $f_{\rm inf}\, f_\star^{-2/3}$ varies slowly with mass, this in turn enforces the well known Fall relationship $j_\star\propto M_\star^{0.5-0.6}$.

\begin{figure}
\centering
\includegraphics[height=9cm]{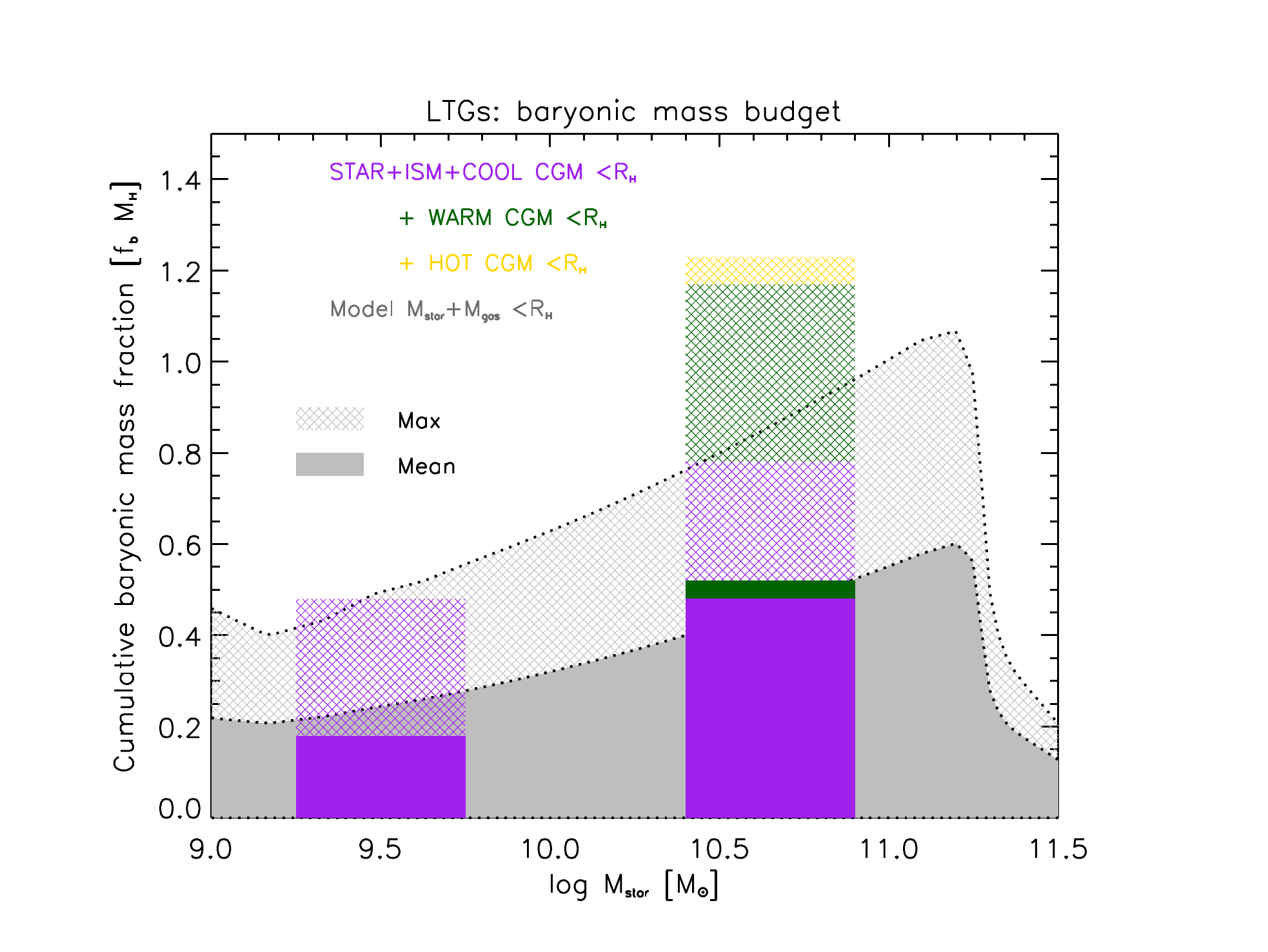}\\\includegraphics[height=9cm]{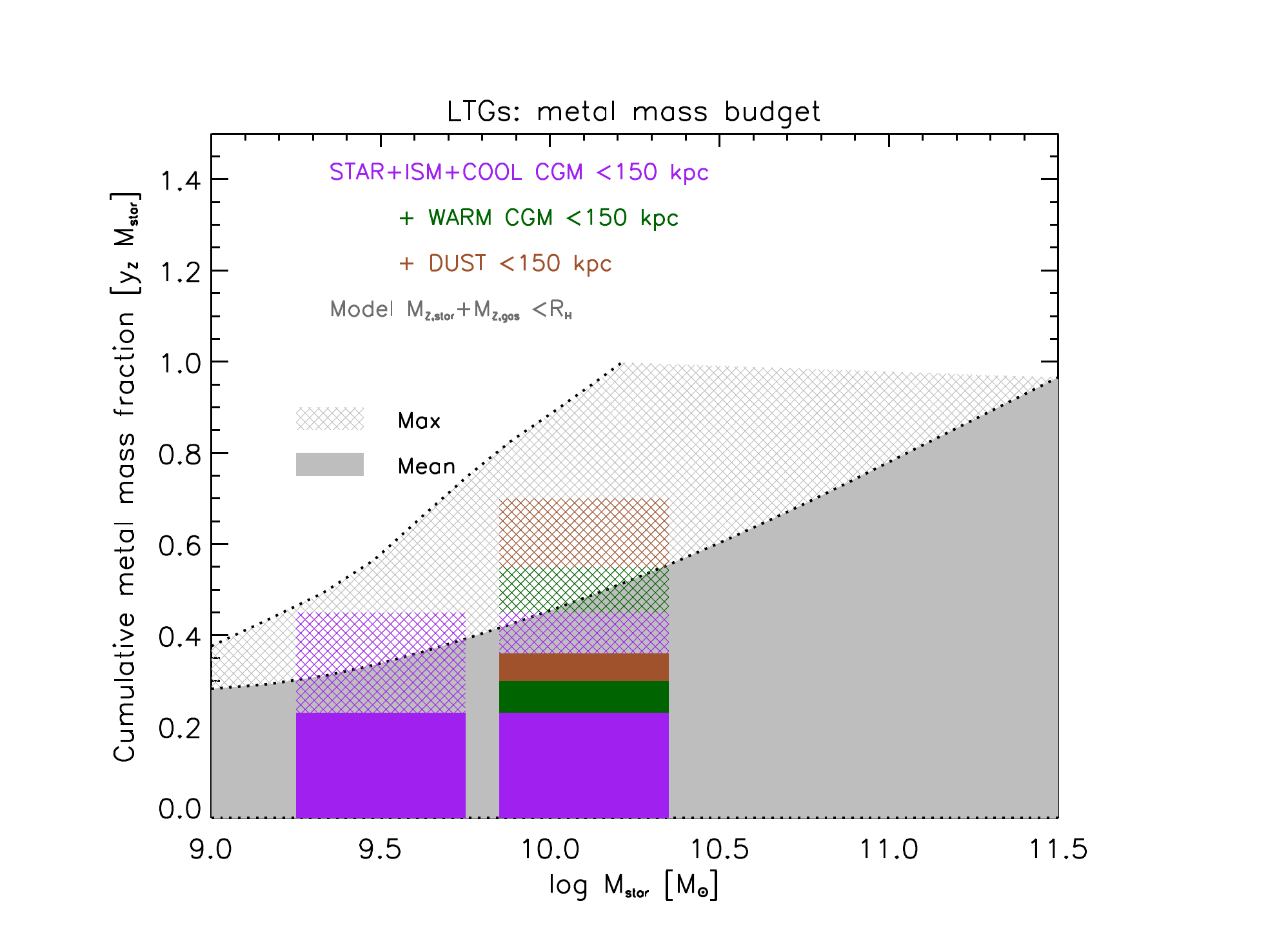}
\caption{Cumulative mass fraction in units of $f_{\rm b}\, M_{\rm H}$ (top panel) and cumulative metal mass fraction in units of $y_Z\, M_\star$ (bottom panel) as a function of the stellar mass $M_\star$. Data collected by Tumlinson et al. (2017; see also Peeples et al. 2014, Werk et al. 2014, Bordoloi et al. 2014) refer to stars+ISM+cold CGM (purple), +warm CGM (green), + hot CGM (yellow, only top panel), + dust (brown, only bottom panel); these are confronted to our results for $M_\star+M_{\rm cold}+M_{\rm inf}$ (grey). In both panels fully colored areas refer to mean values and hatched areas to upper limits (for the model, maximal values are associated to the average over different formation redshifts).}\label{fig|cgm}
\end{figure}

In Fig.~\ref{fig|cgm} we illustrate the cumulative mass fraction in units of $f_{\rm b}\, M_{\rm H}$ and the cumulative metal mass fraction in units of $y_Z\, M_\star$ as a function of the stellar mass $M_\star$. The data have been collected by Tumlinson et al. (2017; see also Peeples et al. 2014, Werk et al. 2014, Bordoloi et al. 2014) and refer to stars+ISM+cold CGM, adding progressively +warm CGM, + hot CGM (or + metal in dust); fully colored areas refer to mean values, while hatched areas to upper limits.
These data are confronted with the results of our one-zone framework (mean values and variance associated to different formation redshifts) in terms of $M_\star+M_{\rm gas}$, and found to be, within the measurement uncertainties and intrinsic model variance, in pleasing agreement.
In particular, our results indicate, in accord with observational  constraints, that a fraction of the available baryonic mass and metals around $20-40\%$ in low-mass systems and $40-70\%$ in high-mass systems is expected to be located in/around galaxies in the form of stars, cold gas in the disk, and cool CGM; this is crucially obtained by the presence of an efficient wind recycling/galactic fountain, that allow to partially recover mass and metals otherwise removed by stellar feedback. Nonetheless, our results also highlight that a still sizeable amount of baryonic mass and metals is not associated to stars, cold disk gas, and cool medium in/around galaxies (the so-called `missing baryons' and `missing metal' problems). Such mass and metals are expected to be ejected by stellar feedback outside the galactic halo in the intergalactic medium (especially in low-mass systems with the shallower potantial wells), or to be heated up on a high adiabat and hindered from remixing to the cool gas within reasonable timescales; the latter material should then be looked for in the warm or hot CGM (hot corona). However, a detailed investigation of such an issue will on the one hand require a more complex framework able to deal with the multi-phase nature of the CGM, and on the other more robust observational constraints.

In Fig.~\ref{fig|complot} we show more in detail how the results presented in this Section are affected by the efficiency of gas recycling associated to galactic fountains. As a matter of fact, more efficient fountains imply higher star formation efficiency, higher gas fraction, higher stellar and gas metallicity, higher dust masses and lower stellar angular momentum. Such behaviors are mainly originated because an efficient wind recycling cause more pre-enriched gas to be available in the halo for condensation and star formation, so extending the star-formation timescale and boosting the metal enrichment in gas and stars.

\section{Summary}\label{sec|summary}

We have generalized the analytic solutions presented in Pantoni et al. (2019) by including a simple yet effective description of wind recycling and galactic fountains, with the aim of self-consistently investigating the spatially-averaged time evolution of the gas, stellar, metal, and dust content in individual disc-dominated, late-type galaxies (LTGs) hosted within a dark halo of given mass and formation redshift (see Sect.~\ref{sec|ansol} and Figs.~\ref{fig|timevo_z05} and \ref{fig|timevo_z15}).

The basic framework pictures the galaxy as an open, one-zone system comprising three interlinked mass components: infalling halo gas able to cool fast, subject to condensation toward the central regions, and refurnished by wind recycling associated to the galactic fountain; cold disk gas fed by infall and depleted by star formation and stellar feedback (type-II SNe and stellar winds); stellar mass, partially restituted to the cold phase by stars during their evolution. The corresponding metal enrichment history of the cold gas and stellar mass is self-consistently computed using as input the solutions for the evolution of the mass components; the metal equations includes effects of feedback, astration, instantaneous production during star formation. Finally, the dust mass evolution takes into account the formation of grain cores associated to star formation, and of the grain mantles due to accretion onto pre-existing cores; astration of dust by star formation and stellar feedback, and spallation by SN shockwaves are also comprised.

We have then supplemented our solutions with a couple of additional ingredients: (i) specific prescriptions for parameter setting based on observational evidences and physical arguments (see Sect.~\ref{sec|params} and Fig.~\ref{fig|params}); (ii) estimates of the average halo (and associated baryonic gas) mass growth by mergers/accretion from the cosmic web, computed on the basis of state-of-the-art $N-$body simulations and depurated from pseudo-evolution (see Sect.~\ref{sec|halogrowth} and Fig.~\ref{fig|halogrowth}).

We then derive a bunch of fundamental relationships involving spatially-averaged quantities as a function of the observed stellar mass: the star formation efficiency (see Fig.~\ref{fig|fstar} and Fig.~\ref{fig|prob3D}), the gas mass fraction (see Fig.~\ref{fig|fgas}), the gas/stellar metallicity (see Fig.~\ref{fig|Zgas} and Fig.~\ref{fig|Zstar}), the star formation rate (see Fig.~\ref{fig|MS}), the dust mass (see Fig.~\ref{fig|mdust}), the stellar specific angular momentum (see Fig.~\ref{fig|jstar}), and the mass/metal budget (see Fig.~\ref{fig|cgm}). We compare these relationships with the data concerning local LTGs, finding a pleasing overall agreement (see Sect.~\ref{sec|results}). We highlight the crucial relevance of wind recycling and galactic fountains in efficiently refurnishing the gas mass, extending the star-formation timescale, and boosting the metal enrichment in gas and stars. We remark that a major value of our approach is to reproduce, with a unique set of physically motivated parameters, a wealth of observables concerning LTGs.

Further developments of this work may involve an investigation of the radial-dependent properties in disk-dominated galaxies; this would require to consider additional processes like differential/selective winds, stellar mixing, multi-zonal crosstalk, vertical structure, disk viscosity, heating mechanisms, accretion from the hot corona. However, including such effects will make the basic framework presented in Sect.~\ref{sec|ansol} much more complex, and the search for analytic solutions even more challenging (see Appendix). Other related extensions of this work could involve the treatment of disk instabilities to investigate the origin and evolution of bars and (pseudo-)bulges in LTGs, and a comparison of the angular momentum evolution in ETGs and LTGs.

To sum up, the analytic solutions provided here are based on an idealized albeit non-trivial descriptions of the main physical processes regulating galaxy formation on a spatially-average ground. Yet, our solutions are simple enough to easily disentangle the role of the main physical processes at work, to allow a quick exploration of the parameter space, and to make transparent predictions on spatially-averaged quantities. As such, they can provide a basis for improving the (subgrid) physical recipes presently implemented in more sophisticated semi-analytic models and numerical simulations, and can offer a benchmark for interpreting and forecasting current and future spatially-averaged observations of local and higher redshift LTGs.

\begin{acknowledgements}
We acknowledge the anonymous referee for constructive comments. We warmly thank A. Bressan, L. Zanisi, P. Salucci and F. Shankar for useful discussions. This work has been partially supported by PRIN MIUR 2017 prot. 20173ML3WW 002, `Opening the ALMA window on the cosmic evolution of
gas, stars and supermassive black holes'. A.L. acknowledges the
MIUR grant `Finanziamento annuale individuale attivit\'a base
di ricerca' and the EU H2020-MSCA-ITN-2019 Project 860744
`BiD4BEST: Big Data applications for Black hole Evolution STudies'.
\end{acknowledgements}

\begin{appendix}

\section{Appendix: analytic solutions with steady accretion rate}\label{sec|appendix}

For completeness, in this Appendix we provide additional analytic solutions (for the mass components) that require minimal extension of the basic framework presented in Sect.~\ref{sec|ansol}. Specifically, we include an additional term $\dot M_{\rm acc}$ describing gas accretion; this may render, e.g., a cold gas flow into the halo from cosmological scales (e.g., Bouche et al. 2010; Dave et al.
2012; Lilly et al. 2013; Dekel \& Mandelker 2014) or a fountain-driven gas condensation from the hot galactic corona (e.g., see Marinacci et al. 2010, 2011; Marasco et al. 2012; Fraternali et al. 2015; Pezzulli \& Fraternali 2016). To retain analytic solutions with simple expressions, we assume a steady accretion rate $\dot M_{\rm acc}(\tau)\simeq$ const. The relevant Eqs.~(\ref{eq|basics}) are just modified as
\begin{equation}\label{eq|appbasics}
\left\{
\begin{aligned}
\dot M_{\rm inf} &= -\frac{M_{\rm inf}}{\tau_{\rm cond}}+\alpha_{\rm GF}\, \epsilon_{\rm out}\, \frac{M_{\rm cold}}{\tau_\star}+\dot M_{\rm acc}~,\\
\\
\dot M _{\rm cold} &= \frac{M_{\rm inf}}{\tau_{\rm cond}} - (1-\mathcal{R})\,\frac{M_{\rm cold}}{\tau_\star}-\epsilon_{\rm out}\, \frac{M_{\rm cold}}{\tau_\star}~,\\
\\
\dot M_\star &= (1-\mathcal{R})\,\frac{M_{\rm cold}}{\tau_\star}~,\\
\end{aligned}
\right.
\end{equation}
with initial condition $M_{\rm inf}(0)=f_{\rm inf}\, M_{\rm b}$ and $M_{\rm cold}(0)=M_\star(0)=0$.

\begin{figure}
\centering
\includegraphics[height=0.8\textwidth]{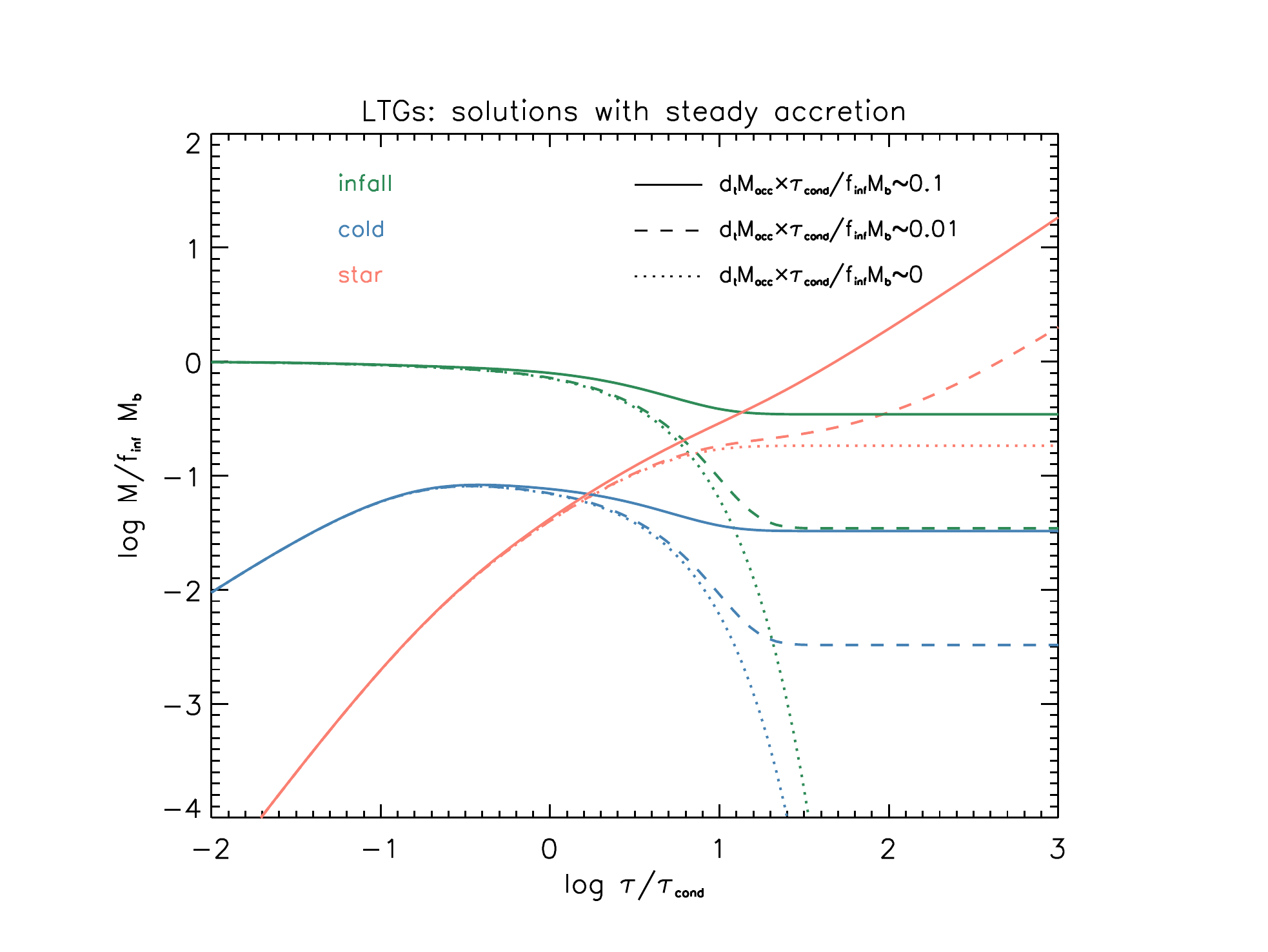}
\caption{Evolution of the mass components for the analytic solutions with steady accretion presented in the Appendix; masses are normalized to the available baryons at halo formation $f_{\rm inf}\, M_{\rm b}$ and the galaxy age $\tau$ is normalized to $\tau_{\rm cond}$. Green lines correspond to the infalling gas mass $M_{\rm inf}$, blue lines to the cold star-forming gas mass $M_{\rm cold}$, and orange lines to the stellar mass $M_\star$. Linestyles refer to different values of $\dot M_{\rm acc}\, \tau_{\rm cond}/f_{\rm inf}\, M_{\rm b}\approx 0.1$ (solid), $0.01$ (dashed) and $0$ (dotted; e.g., the basic solutions of the main text).}\label{fig|appendix}
\end{figure}

In fact, when $\dot M_{\rm acc}(\tau)$ is constant with respect to the galactic age $\tau$, a particular solution of the inhomogeneous equation is easily found as $M_{\rm inf}\simeq  s\gamma\,\dot M_{\rm acc}\, \tau_{\rm cond}/\lambda_+\lambda_-$ and $M_{\rm cold}\simeq \dot M_{\rm acc}\, \tau_{\rm cond}/\lambda_+\lambda_-$. Thus the overall solution writes down
\begin{align}
\left\{
\begin{aligned}
\\
M_{\mathrm{inf}}(\tau) &= \frac{f_{\rm inf}\,M_{\mathrm{b}}}{\lambda_+-\lambda_-}\, \left[(\lambda_+-1)\,\,e^{-\lambda_-\,x} +(1-\lambda_-)\,e^{-\lambda_+\,x}\right]+~,\\
&+ {s\gamma\,\dot M_{\rm acc}\, \tau_{\rm cond}\over \lambda_+\lambda_-}\,\left[1-{\lambda_+-1\over \lambda_+-\lambda_-}\,{\lambda_+\over s\gamma}\,e^{-\lambda_-\,x}+{   1-\lambda_-\over \lambda_+-\lambda_-}\,{\lambda_-\over s\gamma}\,e^{-\lambda_+\,x}\right]
\\
\\
M_{\mathrm{cold}}(\tau) &= \frac{f_{\rm inf}\,M_{\mathrm{b}}}{\lambda_+-\lambda_-}\, \left[e^{-\lambda_-\,x}-e^{-\lambda_+\,x}\right]+\\
&+ {\dot M_{\rm acc}\, \tau_{\rm cond}\over \lambda_+\lambda_-}\,\left[1-{\lambda_+\over \lambda_+-\lambda_-}\,e^{-\lambda_-\,x}+{\lambda_-\over \lambda_+-\lambda_-}\,e^{-\lambda_+\,x}\right]~,\\
\\
M_{\star}(\tau) &= (1-\mathcal{R})\, \frac{s\,f_{\rm inf}\,M_{\mathrm{b}}}{\lambda_+-\lambda_-}\, \left[\frac{1-e^{-\lambda_-\,x}}{\lambda_-}-\frac{1-e^{-\lambda_+\,x}}{\lambda_+}\right]+\\
& + {\dot M_{\rm acc}\, \tau_{\rm cond}\over \lambda_+\lambda_-}\,\left[x-{\lambda_+\over \lambda_-}{1-e^{-\lambda_-\,x}\over \lambda_+-\lambda_-}+{\lambda_-\over \lambda_+}{1-e^{-\lambda_+\,x}\over \lambda_+-\lambda_-}\right]~
\end{aligned}
\right.\\
\nonumber
\end{align}
We illustrate in Fig.~\ref{fig|appendix} the evolution of the mass components for different values of the accretion rate $\dot M_{\rm acc}$ (actually we vary the normalized quantity $\dot M_{\rm acc}\, \tau_{\rm cond}/f_{\rm inf}\, M_{\rm b}$); the parameters $\epsilon_{\rm out}\approx 10$, $s\approx 1$, and $\alpha_{\rm GF}\approx 0.75$ have been set to typical values for Milky-Way like galaxies. All in all, the early evolution is similar to that without steady accretion, while at late times the latter takes over and causes progressive saturation of the cold and infall mass and a linear growth of the stellar one. Plainly, if a switch off of the accretion after some time takes place, the exhaustion of the gas masses and the saturation of the stellar one will be enforced again, but with a substantial delay with respect to our basic model without accretion.

\end{appendix}

\newpage

\newpage

\newpage

\newpage

\newpage

\newpage

\newpage

\newpage

\newpage

\newpage

\newpage

\newpage

\newpage

\newpage

\newpage

\end{document}